%% file: main.tex
\documentclass[a4paper,11pt,notoc]{article}
\usepackage{jheppub} % for details on the use of the package, please see the JINST-author-manual
\usepackage{float}
\usepackage{subfig}
\usepackage{ulem}
\usepackage[table]{xcolor}
\usepackage[compat=1.1.0]{tikz-feynman}
\newcommand{\be}{\begin{equation}}
\newcommand{\ee}{\end{equation}}
\def\bsp#1\esp{\begin{split}#1\end{split}}
\def\bpm{\begin{pmatrix}}
\def\epm{\end{pmatrix}}

\preprint{LAPTH-054/24}
\title{\boldmath Phenomenology of a singlet-doublet-triplet scotogenic framework}

\author[a]{Ugo de Noyers}
\author[b]{~~Maud Sarazin}
\author[a]{~~Bj\"orn Herrmann}

\affiliation[a]{LAPTh, Univ.\ Savoie Mont Blanc, CNRS, F-74000 Annecy, France}
\affiliation[b]{Physics Department, Universidad de los Andes, Bogotá, Colombia}

\emailAdd{denoyers@lapth.cnrs.fr}
\emailAdd{m.sarazin@uniandes.edu.co}
\emailAdd{herrmann@lapth.cnrs.fr}

\abstract{We present an extensive phenomenological study of a scotogenic framework including a scalar singlet, a scalar doublet, a fermionic doublet, and two generations of a fermionic triplet, allowing to provide three non-zero neutrino masses and three viable dark matter candidates. Using a Markov Chain Monte Carlo numerical technique, we probe the parameter space of the model in view of numerous constraints stemming from dark matter, the neutrino sector, and lepton-flavour violating transitions. The dark matter mass distribution of phenomenologically viable scenarios exhibits three very predictive peaks at 550 GeV, 1080 GeV, and 2300 GeV. In all viable parameter regions, the observed relic density is achieved through dominant co-annihilations. Future experiments will be able to probe a large number of the currently viable parameter space, while certain parameter configurations promise interesting collider signatures. For related studies, five typical parameter configurations are given in SLHA format.}


\begin{document}
\notoc
\maketitle
\flushbottom

% =================================================================================
\newpage
\input{intro}
\input{model}
\input{spectrum}
\input{constraints}
\input{results}
\input{conclusion}

% =================================================================================
\acknowledgments

The authors would like to thank W.~Porod for useful discussions, particularly concerning the use of {\tt SARAH 4.15.1} and {\tt SPheno 4.0.5}, and M.~Dubau for helpful discussions concerning the efficient treatment of the data obtained from the MCMC analysis. The work of U.\,d.\,N.\ is funded by a Ph.D.\ grant of the French Ministry for Education and Research. This work is supported by {\it Investissements d’avenir}, Labex ENIGMASS, contrat ANR-11-LABX-0012. M.\,S.\ thanks the constant and enduring financial support received for this project from the physics department and the faculty of science at Universidad de Los Andes (Bogot\'a, Colombia). The plots presented in this article have been obtained using {\tt MatPlotLib} \cite{MatPlotLib}.

% =================================================================================
\appendix
\input{appendix}

% =================================================================================
\bibliographystyle{JHEP}
\bibliography{biblio}

% =================================================================================
\end{document}

%% file: intro.tex
%=====================================
\section{Introduction}
\label{Sec:Intro}
%=====================================

The Standard Model (SM) of particle physics is the prevailing theory that describes three of the fundamental interactions and classifies all the observed particles. It has successfully predicted a wide range of phenomena and its predictions have been tested to very good precision in numerous experiments. Despite these successes, the SM leaves open questions. Notably, in addition to shortcomings of theoretical nature, it lacks a candidate for dark matter (DM) \cite{Planck2018}, and predicts that neutrinos are massless, which contradicts experimental evidence of neutrino oscillations \cite{NuFit2020}.

To address these open questions, it is common extend the SM and explore beyond the Standard Model (BSM) theories. One key area of investigation is the origin of neutrino masses. In the SM, unlike other fermions, neutrinos do not acquire mass through the Higgs mechanism. The true nature of neutrinos—whether they are Dirac or Majorana particles—remains unresolved, allowing for different mass generation mechanisms in BSM models. One possibility is the Seesaw mechanism  \cite{Minkowski:1977sc, Yanagida:1980xy, PhysRevLett.56.561}, considering neutrinos as Dirac particles and including heavy ``right-handed'' neutrinos. In this work, we consider the possibility of Majorana neutrinos acquiring loop-induced mass terms \cite{Ma:2006km}.

DM constitutes another major unresolved issue in the SM. Astrophysical and cosmological observations strongly suggest the presence of DM, which makes up about $27\%$ of the universe’s energy density, yet it remains undetected in laboratory experiments. A promising DM candidate is the Weakly Interacting Massive Particle (WIMP), which is massive, electrically neutral, and stable. WIMPs can account for the observed DM relic density through a process known as thermal freeze-out \cite{Planck2018}.

In this paper, we focus on a framework that addresses both the generation of neutrino masses and the nature of DM. The two aspects are linked by the fact that the neutrino masses are induced through loops containing states from the dark sector, leading to what is known as a ``scotogenic'' model, originally proposed in Ref.\ \cite{Ma:2006km}. This class of models has gained considerable attention over the last decade \cite{Toma:2013zsa, Vicente:2014wga, Fraser:2014yha, Baumholzer:2019twf, Esch2018, Sarazin:2021nwo, Alvarez:2023dzz,Betancur:2017dhy,Betancur:2018xtj} due to their ability to simultaneously address multiple open questions in particle physics. Depending on the exact field content, the scotogenic scenario may be realized based on different  topologies for radiative neutrino mass generation \cite{Restrepo2013}.

A consequence of the chosen field content is the presence of lepton-flavour violating effects, occurring at energy scales that are reachable at current experiments. Such transitions are forbidden in the SM and experimentally well constrained \cite{ParticleDataGroup:2022pth}, especially for $\mu-e$ transitions. Consequently, they put strong restrictions on the physically viable parameter space of the considered framework. Additional constraints stem from the observed mass of the Higgs boson \cite{ParticleDataGroup:2022pth} and the cold DM relic density, which is determined to very high precision in the $\Lambda$CDM cosmological model \cite{Planck2018}.

In the present paper, we focus on a scotogenic framework, where the SM is extended by a scalar singlet, a scalar doublet, a fermionic doublet, and two generations of a fermionic triplet. The proposed framework provides three possible viable candidates for DM, while generating three non-zero neutrino masses. $CP$-violation in the lepton sector is included through two Majorana phases in the Pontecorvo-Maki-Nakagawa-Sakata (PMNS) matrix \cite{Pontecorvo1957b, Maki1962}. First studies of triplet models have been published in Refs.\ \cite{Betancur:2017dhy, Betancur:2018xtj}. In the present study, we make use of a Markov Chain Monte Carlo (MCMC) \cite{Markov1971, Metropolis1953, Hastings1970} algorithm to efficiently confront the full parameter space to the experimental constraints in order to obtain the viable regions of parameter space. 

The present analysis is an extension of and complementary to the finding presented in Refs.\ \cite{Sarazin:2021nwo, Alvarez:2023dzz}. Ref.\ \cite{Sarazin:2021nwo} is based on a somewhat simpler ``topology'', featuring a singlet and a doublet in the fermion sector, allowing for the generation of only two non-zero neutrino masses. In order to obtain a third non-vanishing mass, additional degrees of freedom are required. While the analysis of Ref.\ \cite{Alvarez:2023dzz} includes an additional fermionic singlet, in the present study we replace the fermionic singlet by two generations of a fermionic triplet. In addition to obtaining three non-zero neutrino masses, this is an opportunity to study the phenomenology associated to the triplets, which has, to our knowledge, not been studied extensively throughout the available parameter space.

The present paper is organised as follows: We present the considered model in Sec.\ \ref{Sec:Model} and discuss the physical mass spectrum in Sec.\ \ref{Sec:MassSpectrum}. Sec.\ \ref{Sec:Constraints} is devoted to the discussion of the considered constraints and the computational setup. We discuss our results in Sec.\ \ref{Sec:Results}, before concluding in Sec.\ \ref{Sec:Conclusion}. Additional technical details can be found in the Appendices.

%% file: model.tex
%=====================================
\section{The model Lagrangian}
\label{Sec:Model}
%=====================================

We consider a scotogenic framework, where the SM gauge groupe is extended by a $\mathbb{Z}_2$ symmetry, under which all SM states are even, while all additional states are odd. The field content is extended by a real scalar singlet $S$, a scalar doublet $\eta$, two generations of the same fermion triplet $\Sigma_1$ and $\Sigma_2$, and two Weyl fermion doublets $\Psi_1$ and $\Psi_2$, which are in fact respectively the left and the right components of a Dirac doublet. These extra fields do not carry colour charge, i.e.\ they are $SU(3)_C$ singlets. A summary of the respective representations under $SU(2)_L \times U(1)_Y$ is given in Table \ref{Tab:T12GQuantumNumbers}. This specific scotogenic framework is labelled ``T1-2G'' according to the classification given in Ref.\ \cite{Restrepo2013}.

\begin{table}
    \centering
    \begin{tabular}{|c||c|c|c|c||c|c|}
    \hline
            \     & ~$\Psi_{1}$~ & ~$\Psi_2$~ & ~$\Sigma_1$~ & ~$\Sigma_2$~ & ~$\eta$~ & ~$S$~ \\
            \hline
            \hline
      $SU(2)_L$   &  $\mathbf{2}$ & $\mathbf{2}$ & $\mathbf{3}$ & $\mathbf{3}$ & $\mathbf{2}$ & $\mathbf{1}$ \\
      \hline
        $U(1)_Y$  & -1 & 1 & 0 & 0 & 1 & 0 \\
        \hline
    \end{tabular}
    \caption{The additional fields of the so-called "T1-2G" scotogenic model.}
    \label{Tab:T12GQuantumNumbers}
\end{table}

% --------------------------------------------------------------
\subsection{The scalar sector}
\label{Subsec:Scalars}
% --------------------------------------------------------------

The scalar sector of the model under consideration consists of the SM Higgs doublet $H$, and two additional scalar fields: the singlet $S$ and the doublet $\eta$. Upon electroweak symmetry breaking (EWSB), only $H$ acquires a vacuum expectation value $v = \sqrt{2} \langle H \rangle \approx 246$ GeV. In component notation, the two doublets are then written as
\begin{equation}
    H ~=~ \begin{pmatrix} G^+ \\ \frac{1}{\sqrt{2}} \big( v + h^0 + i G^0 \big) \end{pmatrix}, \qquad
    \eta ~=~ \begin{pmatrix} \eta^+ \\ \frac{1}{\sqrt{2}} \big( \eta^0 + i A^0 \big) \end{pmatrix} \,.
    \label{Eq:ScalarDoublets}
\end{equation}
Here, $G^0$ and $G^+$ are the Goldstone bosons and $h^0$ the physical Higgs boson. The new doublet $\eta$ contains a charged scalar $\eta^+$, a neutral $CP$-even scalar $\eta^0$, and a neutral $CP$-odd pseudoscalar $A^0$. Note that, as we suppose the $\mathbb{Z}_2$ symmetry to be unbroken, the doublet $\eta$ does not acquire a vacuum expectation value. Thus, the scalar sector is similar to the one discussed in the ``T1-2A'' framework in Refs.\ \cite{Sarazin:2021nwo, Alvarez:2023dzz}.

The Lagrangian of the scalar sector of this model is given by 
\begin{equation}
    \begin{split}
        - \mathcal{L}_{\text{scalar}} ~&=~ M^2_H \vert H \vert^2 + \lambda_H \vert H \vert^4 
        + \frac{1}{2} M^2_S S^2 + \frac{1}{2} \lambda_{4S} S^4 + M^2_{\eta} \vert \eta \vert^2 + \lambda_{4\eta} \vert \eta \vert^4 \\ 
        &~~~~+~ \frac{1}{2} \lambda_S S^2 \vert H \vert^2  + \frac{1}{2} \lambda_{S \eta} S^2 \vert \eta \vert^2 + \lambda_{\eta} \vert \eta \vert^2 \vert H \vert^2 +  \lambda^{'}_{\eta} \vert \eta^{\dagger} H \vert^2 \\ 
        &~~~~+~ \frac{1}{2} \lambda^{''}_{\eta} \Big( \big( \eta^{\dagger} H  \big)^2 + \text{h.c.} \Big) + \kappa \left( S \eta^{\dagger} H + \text{h.c.} \right) \,.
    \end{split}
    \label{Eq:ScalarLagrangian}
\end{equation}
We suppose all scalar couplings to be real. The first two terms of the second line are the SM terms related to the Higgs doublet $H$. At tree level, after EWSB, the usual minimization relation in the Higgs sector, 
\begin{equation}
    m^2_{h^0} ~=~ -2 M_H^2 ~=~ 2 \lambda_H v^2 \,,
    \label{Eq:HiggsMass}
\end{equation}
allows to eliminate the free mass parameter $M^2_H$ in favour of the Higgs self-coupling $\lambda_H$. Imposing $m_{h^0} \approx 125 \; \text{GeV}$ leads to a tree-level value of $\lambda_H \approx 0.13$.

The remaining terms of the second line are the mass terms and self-couplings of the new singlet and new doublet. The terms in the third and last line of Eq.\ \eqref{Eq:ScalarLagrangian} include all possible couplings between the Higgs doublet, the new doublet and the new singlet. Mixing between the singlet and the doublet is induced by the trilinear coupling $\kappa$.

The physical mass spectrum and mixing patterns of the the scalar sector will be discussed in Sec.\ \ref{Subsec:ScalarMasses}.

% --------------------------------------------------------------
\subsection{The fermion sector}
\label{Subsec:Fermions}
% --------------------------------------------------------------

In addition to the SM fermions, the scotogenic "T1-2G" framework under consideration contains a Dirac doublet $\Psi$ and two generations $\Sigma_{1,2}$ of a Majorana triplet. The doublet can be written as 
\begin{equation}
    \Psi ~=~ \begin{pmatrix} \Psi_1\\ \Psi_2 \end{pmatrix} \qquad
    \text{with}\quad \Psi_1 ~=~ \begin{pmatrix} \Psi^0_1 \\ \Psi^-_1 \end{pmatrix} \quad\text{and}\quad
    \Psi_2 ~=~ \begin{pmatrix} -\Psi^+_2 \\ (\Psi^0_2)^{\dag} \end{pmatrix} \,,
    \label{Eq:FermionDoublets}
\end{equation}
where $\Psi_1$ and $\Psi_2$ are two Weyl doublets with opposite hypercharge. The two triplets $\Sigma_{1,2}$ can be represented as
\begin{equation}
    \Sigma_j ~=~ \begin{pmatrix}
        \Sigma^0_j/\sqrt{2} & ~\Sigma^+_j \\ ~\Sigma^-_j & -\Sigma^0_j/\sqrt{2} 
    \end{pmatrix} \qquad \text{with}~ j = 1,2 \,.
    \label{Eq:FermionTriplets}
\end{equation}
The corresponding fermionic Lagrangian reads
\begin{equation}
    -\mathcal{L}_{\text{fermion}} ~=~ M_{\Psi} \Psi_1 \Psi_2  + \frac{1}{2} \sum_{i,j} M_{\Sigma_{ij}} \text{Tr}\big\{ \overline{\Sigma}_i \Sigma_j \big\} + \sum_j y_{1j} \Psi_1 \Sigma_j H + \sum_j y_{2j} \Psi_2 \Sigma_j \tilde{H} + \text{h.c.} \,,
    \label{Eq:FermionLagrangian}
\end{equation}
where we have introduced the notation $\tilde{H} = i \sigma^2 H^c$. The Lagrangian includes mass terms for the doublets, denoted $M_{\Psi}$, and for the triplets, denotes $M_{\Sigma_{ij}}$ with $i,j = 1,2$. We place ourselves in a basis where $M_{\Sigma_{12}} = M_{\Sigma_{21}} = 0$, without a loss of generality. Finally, there are Yukawa couplings $y_{1j}$ and $y_{2j}$ ($j = 1, 2$), inducing mixing between the doublets and one of the triplets.

The physical mass spectrum is discussed in Sec.\ \ref{Subsec:FermionMasses}.

% --------------------------------------------------------------
\subsection{Interaction terms}
\label{Subsec:Interactions}
% --------------------------------------------------------------

The final part of the considered Lagrangian contains the interactions between the SM leptons and the extra fields introduced above. The corresponding interaction terms are given by
\begin{equation}
    -\mathcal{L}_{\text{interaction}} ~=~ g_{\Psi}^{\alpha} L_{\alpha} \Psi_2 S ~+~ g_{\Sigma_j}^{\alpha} \eta \Sigma_j L_{\alpha} ~+~ g_R^{\alpha} e_{\alpha}^c \tilde{\eta} \Psi_1 + \text{h.c.} \, ,
    \label{Eq:InteractionLagrangian}
\end{equation}
where $\tilde{\eta} = i \sigma^2 \eta^c$. The leptons are denoted $L_{\alpha}$ ($\alpha = e, \mu, \tau$) for the left-handed doublets, and $e^c_{\alpha}$ for the right-handed singlets. 

\begin{figure}
    \centering
    \includegraphics[scale=0.37]{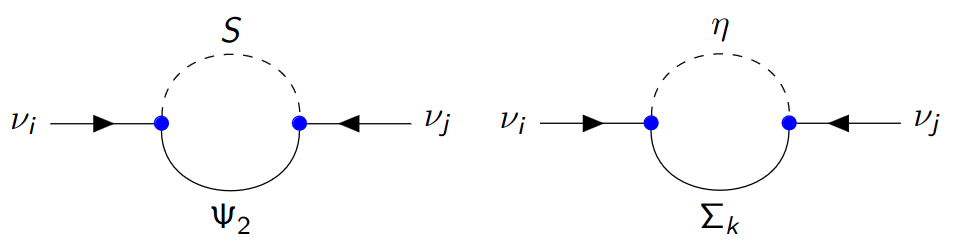}
    \caption{Radiative generation of neutrino masses within the ``T1-2G'' model, depicted in the interaction basis. The blue dots correspond to the couplings introduced in Eq.\ \eqref{Eq:InteractionLagrangian}. The indices $i, j$ run over the three neutrino flavours, $i, j \in \{ e, \mu, \tau \}$, while the index $k$ runs over the two fermionic triplets, $k \in \{ 1,2 \}$.}
    \label{Fig:NeutrinoDiagrams}
\end{figure}

Three-non-zero neutrino masses are radiatively generated through the first two terms of Eq.\ \eqref{Eq:InteractionLagrangian}. Indeed, one can notice that the SM lepton doublet, containing the neutrino fields, is coupled to the new fermionic doublet $\Psi_2$ and the scalar singlet $S$ and to the new fermionic triplets $\Sigma_{1,2}$ and the scalar doublet $\eta$, respectively. The corresponding one-loop diagrams in the interaction basis are depicted in Fig.\ \ref{Fig:NeutrinoDiagrams} and generate the neutrino masses. Note that the last term from Eq.\ \eqref{Eq:InteractionLagrangian} does not contribute to the generation of neutrino masses. Nevertheless, the right-handed coupling $g_R$ will contribute to amplitudes related to LFV processes or DM annihilation and detection.

Finally, one of the objective is to match the current observation of neutrino masses \cite{NuFit2020}. To avoid having to give random values to the coupling, we use the Casas-Ibarra parametrization \cite{CasasIbarra2001}. The following parametrization is used to compute the left-handed couplings in a way where they provide the correct neutrino masses. In order to perform the Casas-Ibarra parametrization, one can write a $3 \times 3 $ matrix that contains the left-handed couplings from Eq. \eqref{Eq:InteractionLagrangian}:
\begin{equation}
    \mathcal{G} ~=~ \begin{pmatrix}
        g_{\Psi}^e & g_{\Psi}^{\mu} & g_{\Psi}^{\tau} \\
        g_{\Sigma_1}^e & g_{\Sigma_1}^{\mu} & g_{\Sigma_1}^{\tau} \\
        g_{\Sigma_2}^e & g_{\Sigma_2}^{\mu} & g_{\Sigma_2}^{\tau}
    \end{pmatrix} \,.
    \label{Eq:CouplingMatrix}
\end{equation}
From Figure \ref{Fig:NeutrinoDiagrams}, we can express the neutrino mass matrix $M_{\nu}$ as
\begin{equation}
    \mathcal{M}_{\nu} ~=~ \mathcal{G}^T \, M_L \, \mathcal{G} \,,
    \label{Eq:NeutrinoMassMatrix}
\end{equation}
where $M_L$ contains the expressions of the loop calculations. The components of $M_L$ depend on the parameters related to the scalar and fermionic sectors, their explicit expressions for the scotogenic framework under consideration are given in App.\ \ref{App:LoopMatrix}.

By making use of the Casas-Ibarra parametrization, assuming a normal mass ordering, the necessary coupling values encompassed in the matrix $\mathcal{G}$ can be computed from the physical neutrino masses and mixing parameters as well as the loop matrix $M_L$. The coupling matrix $\mathcal{G}$ can be related to the other quantities according to
\begin{equation}
    \mathcal{G} ~=~ U_L \, D^{-1/2}_L \, R \, D^{1/2}_{\nu} \, U^{*}_{\text{PMNS}} \,,
    \label{Eq:CouplingMatrixCasasIbarra}
\end{equation}
where the diagonal matrix $D_L$ as well as the rotation matrix $U_L$ stem from diagonalizing the matrix $M_L$,
\begin{equation}
    D_L ~=~ U^t_L \, M_L \, U_L \,.
    \label{Eq:DiagonalLoopMatrix}
\end{equation}
Moreover, $D_{\nu}$ is the diagonal matrix containing the neutrino masses $m_{\nu_1}$, $m_{\nu_2}$ and $m_{\nu_3}$, assuming normal ordering. The PMNS matrix $U_{\text{PMNS}}$ relates the neutrino flavours to their mass eigenstates, after considering that the charged leptons are already in their mass eigenbasis, and is parameterised through three mixing angles, ($\theta_{12}$, $\theta_{13}$ and $\theta_{23}$), the $CP$-violating Dirac phase ($\delta_{CP}$), as well as two $CP$-violating Majorana phases ($\alpha_1$ and $\alpha_2$). The PMNS matrix is detailed in the Appendix \ref{App:CasasIbarra}.

The remaining degrees of freedom are encompassed in the $3 \times 3$ matrix $R$, depending on three complex angles, which are free parameters. A detailed expression of $R$ is given in App.\ \ref{App:CasasIbarra}. 

%% file: spectrum.tex
% ==========================================================================
\section{Physical mass spectrum}
\label{Sec:MassSpectrum}
% ==========================================================================

The physical mass eigenstates of the theory are obtained by diagonalizing the mass matrices arising from the Lagrangians given in Eqs.\ \eqref{Eq:ScalarLagrangian} and \eqref{Eq:FermionLagrangian}. In practice, we use the numerical spectrum calculator {\tt SPheno~4.0.5} \cite{SPheno2003, SPheno2012} to obtain the physical mass spectrum including corrections at the one-loop level. The model has been implemented using the {\tt Mathematica} package {\tt SARAH~4.15.1} \cite{SARAH2010, SARAH2011, SARAH2013, SARAH2014}.

% --------------------------------------------------------------------------
\subsection{Scalar masses and mixing}
\label{Subsec:ScalarMasses}
% --------------------------------------------------------------------------

After EWSB, the neutral scalar states $S$ and $\eta^0$ mix into two $CP$-even neutral mass eigenstates, denoted $\phi_1^0$ and $\phi_2^0$. As we suppose the scalar couplings to be real, there is no mixing with the third neutral component $A^0$, which is $CP$-odd. In the basis $\{ S, \eta^0, A^0 \}$, the tree-level mass matrix is given by
\begin{equation}
    \mathcal{M}^2_{\phi} ~=~ \begin{pmatrix}
        M^2_S + \frac{1}{2} v^2 \lambda_S & v \kappa & 0 \\
        v \kappa & M^2_{\eta} + \frac{1}{2} v^2 \lambda_L & 0 \\
        0 & 0 & M^2_{\eta} + \frac{1}{2} v^2 \lambda_A
    \end{pmatrix} \,,
    \label{Eq:ScalarMatrix}
\end{equation}
where $\lambda_{L,A} = \lambda_{\eta} + \lambda^{'}_{\eta} \pm \lambda^{''}_{\eta}$. The rotation to the physical mass basis is defined as 
\begin{equation}
    \left( \phi_1^0, \phi_2^0, A^0 \right)^t ~=~ U_{\phi} \, \left( S, \eta^0, A^0 \right)^t \,,
    \label{Eq:ScalarMixing}
\end{equation}
where the $CP$-even mass eigenstates are supposed to be ordered such that $m_{\phi^0_1} \leq m_{\phi^0_2}$. Finally, the tree-level mass of the charged scalar eigenstates ($\phi^{\pm}=\eta^{\pm}$) is given by
\begin{equation}
    m^2_{\phi^{\pm}} ~=~ M^2_{\eta} + \frac{1}{2} v^2 \lambda_{\eta} \,.
    \label{Eq:ChargedScalarMassEigenstates}
\end{equation}

\begin{figure}
    \centering
    \includegraphics[scale=0.5]{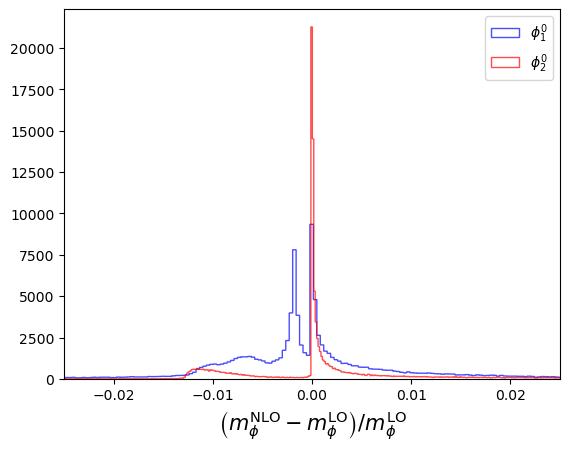}
    \includegraphics[scale=0.5]{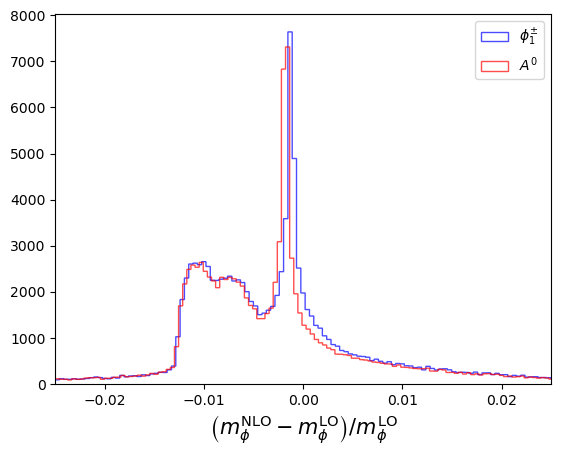}
    \caption{Histograms showing the relative impact of the higher-order corrections for the masses of the $CP$-even ($\phi^0_{1,2}$, left), $CP$-odd ($A^0$, right) and charged ($\phi^{\pm}$, right) scalars for a sample of 117\,000 phenomelogically viable parameter points. For each state, we show the difference of the mass calculated including one-loop corrections ($m_{\phi}^{\text{NLO}}$) and the leading-order mass ($m_{\phi}^{\text{LO}}$) normalised to the latter.}
    \label{Fig:Histo_relative_impact_of_loop_corrections_scalar_sector}
\end{figure}

As stated above, we compute the physical mass spectrum using {\tt SPheno} including radiative corrections at the one-loop level, which typically affect the resulting physical masses by typically not more than 2\%. In Fig.\ \ref{Fig:Histo_relative_impact_of_loop_corrections_scalar_sector}, we show the relative corrections obtained in our study of about 117\,000 parameter points (see Section \ref{Sec:Constraints}). Similar results have been obtained for other scotogenic realisations \cite{Sarazin:2021nwo}. 

% --------------------------------------------------------------------------
\subsection{Fermion masses and mixing}
\label{Subsec:FermionMasses}
% --------------------------------------------------------------------------

\begin{figure}
    \centering
    \includegraphics[scale=0.5]{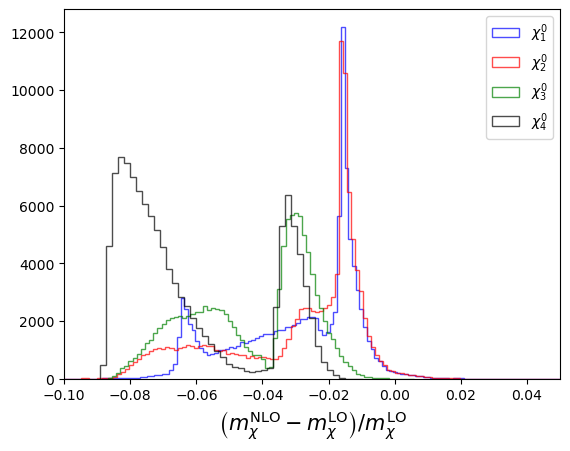}
    \includegraphics[scale=0.5]{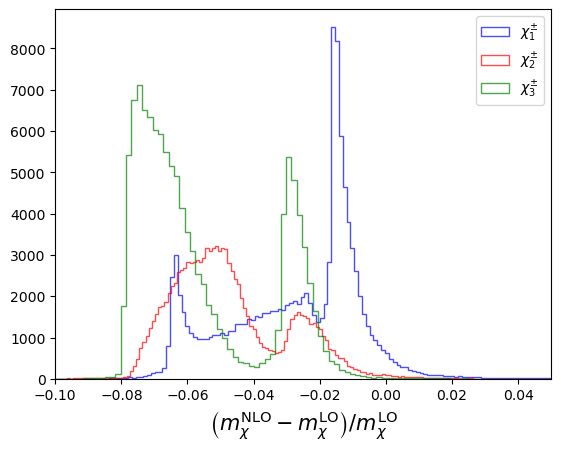}
    \caption{Histograms showing the relative impact of the higher-order corrections for the masses of the $CP$-even ($\chi^0_{1,2,3,4}$, left) and charged ($\chi^{\pm}_{1,2,3}$, right) fermions for a sample of 117\,000 phenomelogically viable parameter points. Same notation as for Fig.\ \ref{Fig:Histo_relative_impact_of_loop_corrections_scalar_sector}.}
    \label{Fig:Histo_relative_impact_of_loop_corrections_fermion_sector}
\end{figure}

In the same way as for the scalars, the neutral components of the fermionic doublets and triplets will mix, leading to four neutral Majorana mass eigenstates, denotes $\chi^0_i$ ($i=1,2,3,4$). In the basis $\{ \Sigma_1^0, \Sigma_2^0, \Psi_1^0, \Psi_2^0 \}$, the tree-level mass matrix is given by
\begin{equation}
    \mathcal{M}_{\chi^0}= \begin{pmatrix}
        M_{\Sigma_{11}} & 0 & - \frac{v y_{11}}{2} & - \frac{v y_{21}}{2} \\
        0 & M_{\Sigma_{22}} & - \frac{v y_{12}}{2} & - \frac{v y_{22}}{2} \\        
        - \frac{v y_{11}}{2} & - \frac{v y_{12}}{2} & 0 & M_{\Psi} \\
        - \frac{v y_{21}}{2} & - \frac{v y_{22}}{2} & M_{\Psi} & 0
    \end{pmatrix} \,.
    \label{Eq:NeutralFermionMatrix}
\end{equation}
Let us recall that $M_{\Sigma_{12}} = M_{\Sigma_{21}} = 0$ without any loss of generality. The above matrix is diagonalized according to 
\begin{equation}
    \left( \chi^0_1, \chi^0_2, \chi^0_3, \chi^0_4 \right)^t ~=~ U_{\chi^0} \, \left( \Sigma_1^0, \Sigma_2^0, \Psi_1^0, \Psi_2^0 \right)^t \,
    \label{Eq:NeutralFermionMixing}
\end{equation}
where the mass eigenstates are ordered such that $m_{\chi_i^0} \leq m_{\chi_j^0}$ for $i<j$. \\

In addition, and contrary to the charged scalars, there is mixing in the charged fermion sector leading to three charged Dirac fermions. In the basis $\big\{\Sigma^+_1, \Sigma^+_2, \Psi^+_1, \Sigma^-_1, \Sigma^-_2, \Psi^-_2 \big\}$, the associated mass term can be written as
\begin{equation}
    -{\cal L}_{\text{fermion}} ~\supset~ \frac{1}{2} \left( \psi^{\pm} \right)^{\dag} \, {\cal M}_{\chi^{\pm}} \, \psi^{\pm} ~+~ \text{h.c.} \,,
\end{equation}
where the mass matrix is given by,
\begin{equation}
    {\cal M}_{\chi^{\pm}} ~=~ \begin{pmatrix} 0 & X^t \\ X & 0 \end{pmatrix}
    \qquad \text{with} \qquad
    X ~=~ \begin{pmatrix}
        M_{\Sigma_{11}} & 0 & - \frac{v y_{11}}{\sqrt{2}} \\
        0 & M_{\Sigma_{22}} & - \frac{v y_{12}}{\sqrt{2}} \\
        - \frac{v y_{21}}{\sqrt{2}} & - \frac{v y_{22}}{\sqrt{2}} & M_{\Psi}
        \end{pmatrix} \,.
    \label{Eq:FermionChargedMatrix}
\end{equation}
The gauge and mass eigenstates are linked through two unitary $3 \times 3$ matrices $U_{\chi^+}$ and $U_{\chi^-}$ according to
\begin{equation}
    \left( \chi^+_1, \chi^+_2, \chi^+_3 \right)^t ~=~ U_{\chi^+} \, \left(\Sigma^+_1, \Sigma^+_2, \Psi^+_1 \right)^t \quad \text{and} \quad \left( \chi^-_1, \chi^-_2, \chi^-_3 \right)^t ~=~ U_{\chi^-} \left(\Sigma^-_1, \Sigma^-_2, \Psi^-_2 \right)^t \,.
    \label{Eq:FermionMassEigenstates}
\end{equation}
These matrices allow to diagonalize the above mass matrix such that 
\begin{equation}
    \text{diag}\big( m_{\chi^{\pm}_1}, m_{\chi^{\pm}_2}, m_{\chi^{\pm}_3} \big) ~=~ U_{\chi^+}^* \, X \, U_{\chi^-} \,,
\end{equation}
with the convention $m_{\chi^{\pm}_i} \leq  m_{\chi^{\pm}_j}$ for $i<j$. This singular value decomposition is the same as the one present in the chargino sector of the Minimal Supersymmetric Standard Model \cite{Martin:1997ns}.

As for the scalar sector, our numerical implementation in {\tt SPheno 4.0.5} takes into account the radiative one-loop corrections to the physical masses. As can be seen in Fig.\ \ref{Fig:Histo_relative_impact_of_loop_corrections_fermion_sector} for the same sample of 117\,000 parameter points as for the scalars shown above, the corrections for the neutral fermion states amount to maximally 7\%. Again, this is similar to the corrections obtained in Ref.\ \cite{Sarazin:2021nwo}.

%% file: constraints.tex
% =====================================
\section{Constraints and computational setup}
\label{Sec:Constraints}
% =====================================

The 33-dimensional parameter space introduced above is subject to a large number of constraints, of theoretical or observational nature, and stemming from different sectors, such as DM, neutrino masses and mixings, scalar sector, and lepton flavour violating processes. We determine the phenomenologically viable parameter regions by means of a MCMC scan \cite{Metropolis1953, Markov1971}, allowing the efficient exploration of the parameter space in view of the various constraints.

In this Section, we first discuss the applied experimental constraints. We then detail the implementation of the MCMC scan together with the calculation of the associated likelihoods.

% --------------------------------------------------------------------------
\subsection{Scalar sector}
\label{Subsec:ScalarConstraints}
% --------------------------------------------------------------------------

In order to assure the existence of a stable vacuum state, the scalar potential needs to conserve the usual ``sombrero shape'' and hence needs to be bounded from below. In the present framework, this translates into a set of inequality conditions for the scalar couplings, given in App.\ \ref{App:ScalarConditions}. In the following analysis, this condition is always imposed.

% --------------------------------------------------------------------------
\subsection{Dark matter}
\label{Subsec:DM}
% --------------------------------------------------------------------------

The model parameter space is constrained by the requirement to meet the observed relic density of DM while not exceeding the spin dependent and spin independent cross-section limits obtained from direct detection searches. In $\Lambda$CDM cosmology, the currently most precise determination of the relic density stems from cosmic microwave background observations made by the Planck mission, leading to the very narrow interval $\Omega_{\rm CDM} h^2 ~=~ 0.1200 \pm 0.0012$ \cite{Planck2018}. We consider the ``freeze-out'' scenario, where the current DM relic density is achieved through annihilation and co-annihilation of the relic particle. In practice, for our MCMC scan, we allow for a slightly larger interval, due to computational \cite{MO2004} and theoretical uncertainties (see, e.g., Ref.\ \cite{Harz:2023llw} and references therein),
\begin{align}
    \Omega_{\rm CDM} h^2 ~=~ 0.1200 \pm 0.0042 \,.
    \label{Eq:RelicDensity}
\end{align}

The detection of DM via direct detection experiments depends on its interaction with baryonic matter, usually Xenon particles because of their stability, characterised by spin-dependent and spin-independent cross sections. In the mass region of our interest, the most stringent cross-section upper limits, obtained from the search for nuclear recoils in a Xenon medium, are provided by the LUX-ZEPLIN (LZ) experiment \cite{LZ:2022lsv}, which is among the latest and most sensitive direct detection experiment.

As we will see in Sec.\ \ref{Sec:Results}, the DM constraints influence mainly the DM mass range and thus the nature of the DM candidate, while certain coupling parameters can also be affected. In our numerical study, we make use of the {\tt micrOMEGAs 5.3.41} code \cite{MO2001, MO2004, MO2007a, MO2007b, MO2013, MO2018} to compute the needed predictions for the DM relic density and the DM-nucleon interaction cross-sections.

% --------------------------------------------------------------------------
\subsection{Lepton flavour and $CP$ violation}
\label{Subsec:LFV}
% --------------------------------------------------------------------------

The couplings $g_{\Psi}^{\alpha}$, $g_{\Sigma}^{\alpha}$ and $g_R^{\alpha}$ ($\alpha=e,\mu,\tau$), introduced in Eq.\ \eqref{Eq:InteractionLagrangian}, give rise to lepton flavour violating (LFV) interactions, mediated at the one-loop level. As no direct evidence for such transitions has been observed, the couplings are required to be reasonably small, in order to satisfy the obtained experimental upper limits on LFV branching ratios and related observables. Let us recall that the values of $g_{\Psi}^{\alpha}$ and $g_{\Sigma}^{\alpha}$ ($\alpha=e,\mu,\tau$) are computed through the Casas-Ibarra parametrization (see Sec.\ \ref{Subsec:Interactions}). As the involved scalar and fermionic fields are supposed to have masses around the TeV scale, they are expected to be numerically small such that they lead to the required relatively low neutrino masses. Consequently, LFV processes will mainly constrain the values of $g_R^{\alpha}$ ($\alpha=e,\mu,\tau$), which do not enter the neutrino mass calculation.

\begin{table}
    \centering
    \begin{tabular}{|c|c|}
        \hline
         \textbf{Observable} & \textbf{Constraint}
         \\
         \hline\hline
        $m_H$ & 125.25 $\pm$ 3.0 GeV
        \\
        \hline
        $\Omega_{\rm CDM}h^2$ & 0.1200 $\pm$ 0.0042
        \\
        \hline
        $\text{BR}(\mu^-$ $\to$ $e^- \gamma)$ & $< 4.2 \times 10^{-13}$
        \\
        \hline
        $\text{BR}(\tau^-$ $\to$ $e^- \gamma)$ & $< 3.3 \times 10^{-8}$
        \\
        \hline
        $\text{BR}(\tau^-$ $\to$ $\mu^- \gamma)$ & $< 4.2 \times 10^{-8}$
        \\
        \hline
        $\text{BR}(\mu^-$ $\to$ $e^- e^+ e^-)$ & $< 1.0 \times 10^{-12}$
        \\
        \hline
        $\text{BR}(\tau^-$ $\to$ $e^- e^+ e^-)$ & $< 2.7 \times 10^{-8}$
        \\
        \hline
        $\text{BR}(\tau^-$ $\to$ $\mu^- \mu^+ \mu^-)$ & $< 2.1 \times 10^{-8}$
        \\
        \hline
        $\text{BR}(\tau^-$ $\to$ $e^- \mu^+ \mu^-)$ & $< 2.7 \times 10^{-8}$
        \\
        \hline
        $\text{BR}(\tau^-$ $\to$ $ \mu^- e^+ e^-)$ & $< 1.8 \times 10^{-8}$
        \\
        \hline
        $\text{BR}(\tau^-$ $\to$ $ \mu^- e^+ \mu^-)$ & $< 1.7 \times 10^{-8}$
        \\
        \hline
        $\text{BR}(\tau^-$ $\to$ $ \mu^+ e^- e^-)$ & $< 1.5 \times 10^{-8}$
        \\
        \hline
        $\text{BR}(\tau^-$ $\to$ $  e^- \pi )$ & $< 8.0 \times 10^{-8}$
        \\
        \hline
    \end{tabular}
    \qquad
    \begin{tabular}{|c|c|}
        \hline
         \textbf{Observable} & \textbf{Constraint}
         \\
         \hline\hline
        $\text{BR}(\tau^-$ $\to$ $ e^- \eta)$ & $< 9.2 \times 10^{-8}$
        \\
        \hline
        $\text{BR}(\tau^-$ $\to$ $ e^- \eta')$ & $< 1.6 \times 10^{-7}$
        \\
        \hline
        $\text{BR}(\tau^-$ $\to$ $  \mu^- \pi )$ & $< 1.1 \times 10^{-7}$
        \\
        \hline
        $\text{BR}(\tau^-$ $\to$ $ \mu^- \eta)$ & $< 6.5 \times 10^{-8}$
        \\
        \hline
        $\text{BR}(\tau^-$ $\to$ $ \mu^- \eta')$ & $< 1.3 \times 10^{-7}$
        \\
        \hline
        $\text{CR}_{\mu \to e}$(Ti) & $< 4.3 \times 10^{-12}$
        \\
        \hline
        $\text{CR}_{\mu \to e}$(Pb) & $< 4.3 \times 10^{-11}$
        \\
        \hline
        $\text{CR}_{\mu \to e}$(Au) & $< 7.0 \times 10^{-13}$
        \\
        \hline
        $\text{BR}(Z^0$ $\to$ $ e^\pm \mu^\mp  )$ & $< 7.5 \times 10^{-7}$
        \\
        \hline
        $\text{BR}(Z^0$ $\to$ $ e^\pm \tau^\mp )$ & $< 5.0\times 10^{-6}$
        \\
        \hline
        $\text{BR}(Z^0$ $\to$ $ \mu^\pm \tau^\mp)$ & $< 6.5 \times 10^{-6}$
        \\
        \hline
        $\text{EDM}_e$ & $< 1.1 \times 10^{-27} $
        \\
        \hline
        $\text{EDM}_{\mu}$ & $< 1.8 \times 10^{-19}$
        \\
        \hline
        $\text{EDM}_{\tau}$ & $< 1.15 \times 10^{-15} $
        \\
        \hline
    \end{tabular}
    \caption{Constraints from the Higgs-boson mass ($m_H$), the DM relic density ($\Omega_{\rm CDM}h^2$), and numerous LFV and $CP$-violating observables (remaining entries) \cite{ParticleDataGroup:2022pth} as implemented in our MCMC analysis.}
    \label{Tab:constraints}
\end{table}

The most stringent constraints stem from $\mu-e$ transitions, such as the branching ratio of the decay $\mu \to e\gamma$, measured by the MEG experiment \cite{FromPDG_TheMEG:2016wtm}, the decay $\mu \to 3e$, measured by Mu3e \cite{Blondel:2013ia}, and the $\mu-e$ conversion rate in nuclei, e.g.\ in Au, measured by SINDRUM II  \cite{SINDRUMII:2006dvw}, Mu2e, COMET, and DeeME \cite{Rule:2024kjo, Calibbi:2017uvl, Natori:2014yba}. Let us note that, while the current sensitivity already constitutes a rather important constraint, it is expected to be improved with the upcoming MEGII experiment \cite{Meucci:2022qbh} as well as improvements planned for COMET to reach CR$_{\mu \to e}({\rm Al}) < 10^{-17}$ \cite{COMET2024, Moritsu:2022lem, deGouvea:2013zba}. 

In our MCMC study, we include all available observables related to lepton flavour violating transitions, as summarised in Table \ref{Tab:constraints}. In addition, we include the constraints on the electric dipole moments of the electron, muon, and tau lepton, as indicated in Table \ref{Tab:constraints} \cite{ParticleDataGroup:2022pth}. We make use of {\tt SPheno 4.0.5} to numerically compute the associated predicted values for a given parameter point. Let us already note that the electric dipole moments (EDMs) do not have a significant impact within our study as all additional $CP$-violating effects with respect to the Standard Model turn out to be negligible.

% --------------------------------------------------------------------------
\subsection{Computational setup}
\label{Subsec:Setup}
% --------------------------------------------------------------------------

In the spirit of previous analyses of the ``T1-2A'' scotogenic framework \cite{Sarazin:2021nwo, Alvarez:2023dzz}, we employ a MCMC \cite{Markov1971} code, based on the Metropolis-Hastings \cite{Metropolis1953} algorithm, allowing to efficiently explore the parameter space in view of the imposed constraints listed in Table \ref{Tab:constraints}. In addition, we take into account the DM direct detection limits from LUX-ZEPLIN (LZ) \cite{LZ:2022lsv}. Note that for the DM relic density $\Omega_{\text{CDM}}h^2$ and the Higgs boson mass $m_H$, the theory uncertainties are larger than the experimental ones, and consequently, the theory uncertainties will be applied. Since we require a Weakly Interactive Massive Particle (WIMP) candidate, we ensure also that our lightest $\mathbb{Z}_2$-odd particle is electrically neutral. Finally, we take into account the vacuum stability conditions mentioned in Sec.\ \ref{Subsec:Scalars} and detailed in App.\ \ref{App:ScalarConditions}.

The MCMC technique allows to explore, in an iterative procedure, the parameter space of the model. This exploration is conditioned by the computation of a likelihood value associated to each explored parameter set in view of the imposed constraints.

Our MCMC scan runs over 33 free parameters, including couplings and mass terms coming from the Lagrangian, the lightest neutrino mass, as well as the complex angles of the rotation matrix $R$ appearing in Eq.\ \eqref{Eq:CouplingMatrixCasasIbarra}. We assume a normal ordering for the neutrino masses, and we use the latest NuFit data \cite{NuFit2020, NuFit2019, NuFit2018} to compute $m_{\nu_2}$ and $m_{\nu_3}$. The mass ranges are chosen to be in the reach of the high luminosity collider LHC (HL-LHC). The couplings from the scalar sector have to satisfy the conditions on the vacuum stability and give a scalar potential that is bounded from below.

For the sake of efficiency in probing the maximum of the parameter space, we chose to use a logarithmic scale for all parameters, and we assign a possible sign on a random basis. This scan is performed, over the parameter ranges specified in Table \ref{Tab:setup}, yielding a total of $117 \, 000$ viable points contained in chains of 320 points each. The first 80 points (``burn-in length'') of each chain have been removed in order to keep only the points for which the likelihood value is reasonably large. 

\begin{table}
    \centering
    \begin{tabular}{|c|c|}
        \hline
         \textbf{Parameter} & \textbf{Interval}
         \\
         \hline\hline
        $\lambda_H$ & $\left[0.1, 0.4 \right]$
        \\
        \hline
        $\lambda_{S}, \lambda_{S \eta}$ & $\left[-1, 1 \right]$
        \\
        \hline
        $\lambda_{4S}, \lambda_{4 \eta}$ & $\left[10^{-7}, 1 \right]$
        \\
        \hline
        $\lambda_{\eta}, \lambda^{'}_{\eta}, \lambda^{''}_{\eta}$ & $\left[-1, 1 \right]$
        \\
        \hline
        $\kappa$ & $\left[-10^3, 10^3 \right]$
        \\
        \hline
        $M^2_S, M^2_{\eta}$ & $\left[2 \times 10^{5}, 16 \times 10^{6} \right]$
        \\
        \hline
        $M_{\Psi}, M_{\Sigma_{jj}}$ & $\left[6 \times 10^{2}, 4 \times 10^{3} \right]$
        \\
        \hline
        $y_{ij}$ & $\left[-1, 1 \right]$ 
        \\
        \hline
        $g_R^{\alpha} $ & $\left[-1, 1 \right]$ 
        \\
        \hline
        $m_{\nu_1}$ & $\left[10^{-32}, 10^{-10} \right]$ 
        \\
        \hline
    \end{tabular}
    \qquad
    \begin{tabular}{|c|c|}
        \hline
         \textbf{Parameter} & \textbf{Interval}
         \\
         \hline \hline
        $\Delta m^2_{21}$ & $\left[ 7.01, 7.82 \right]  \times 10^{-23}$
        \\
        \hline
        $\Delta m^2_{31}$ & $\left[ 2.457, 2.567\right] \times 10^{-21}$
        \\
        \hline
        $\mathfrak{Re}(r_k)$ & $\left[ -1, 1 \right]$
        \\
        \hline
        $\mathfrak{Im}(r_k)$ & $\left[ -1, 1 \right]$
        \\
        \hline
        $\alpha_{1}$ & $\left[ 0, 180 \right]$
        \\
        \hline
        $\alpha_{2}$ & $\left[ 0, 180 \right]$
        \\
        \hline 
        $\delta_{CP}$ & $\left[ 147, 281 \right]$
        \\
        \hline        
        $\theta_{12}$ & $\left[ 31.97, 34.91 \right]$
        \\
        \hline
        $\theta_{13}$ & $\left[ 8.3, 8.76 \right]$
        \\
        \hline        
        $\theta_{23}$ & $\left[ 46.5, 51.2 \right]$
        \\
        \hline
    \end{tabular}
    \caption{Parameters that are varied in our MCMC scan and associated parameter ranges. Left: Input parameters of the scalar and fermion sectors for the MCMC scan. Right: Parameters of the neutrino sector involved in the Casas-Ibarra parametrization. Indices run such that $i,j \in \{1,2\}$, $k \in \{ 1, 2, 3\}$, and $\alpha \in \{ e, \mu, \tau \}$.}
    \label{Tab:setup}
\end{table}

To perform the MCMC scan we implement the model in {\tt SARAH 4.15.1} \cite{SARAH2010, SARAH2011, SARAH2013, SARAH2014} and generate the numerical module for {\tt SPheno 4.0.5}  \cite{SPheno2003, SPheno2012} allowing to compute the mass spectrum at the one-loop level as well as certain of the considered observables, such as LFV decays and $(g-2)$. We also implement the model in {\tt CALCHEP} \cite{CalcHep_Belyaev:2012qa} to use {\tt micrOMEGAs 5.3.41} \cite{MO2001, MO2004, MO2007a, MO2007b, MO2013, MO2018} for DM related constraints, i.e.\ relic density and direct detection cross-section. The SM parameters are fixed to $G_F = 1.166370 \times 10^{-5} \; \text{GeV}$, $\alpha_S(M_Z) = 0.1187$, $M_Z = 91.1887 \; \text{GeV}$, $m_b(m_b) = 4.18 \;\text{GeV}$, $m_t^{\text{pole}} = 173.5 \; \text{GeV}$ and $m_{\tau}(m_{\tau}) = 1.77669 \; \text{GeV}$. 

By assuming a Gaussian likelihood of uncorrelated observable, the latter is computed as the product of individual likelihoods, with respect to each of the imposed constraints,  
\begin{equation}
    \mathcal{L}_n ~=~ \prod_i \mathcal{L}_i^n\, ,
    \label{Eq:FullSetLikelihood}
\end{equation}
with the index $i$ running over the various constraints and $\mathcal{L}_i^n$ is the individual value likelihood associated to each constraint.

In case of a measured observable, such as the Higgs-boson mass, $m_H$, or the DM relic density, $\Omega_{\text{CDM}} h^2$, experimental intervals have been determined. Consequently, these constraints are implemented assuming a Gaussian profile with the given uncertainty $\sigma_i$. In this case, the individual likelihood of a given observable is computed as 
\begin{equation}
    \ln \mathcal{L}_i^n ~=~ - \frac{\left( \mathcal{O}^n_i - \mathcal{O}^{\text{exp}}_i \right)^2}{2 \sigma^2_i} \,,
    \label{Eq:IndividualLikelihood}
\end{equation}
where $\mathcal{O}_i^n$ is the calculated value of the considered observable for the parameter point $n$, $\mathcal{O}_i^{\text{exp}}$ is the associated experimental value given in Table \ref{Tab:constraints}, and $\sigma$ is the associated uncertainty. In case of experimental upper limits, the likelihood computation is implemented as a step function which is smeared as a single-sided Gaussian with a width of $10\%$ of the value corresponding to the experimental upper limit. In the case where the predicted value $\mathcal{O}_i^n$ is above the upper limit, we compute the likelihood $\mathcal{L}_i^n$ using Eq.\ \eqref{Eq:IndividualLikelihood} with $\mathcal{O}_i^{\text{exp}}$ being the upper limit and $\sigma_i = 0.1 \mathcal{O}_i^{\text{exp}}$. In the opposite case, where $\mathcal{O}_i^n$ is below the upper limit, we set $\mathcal{L}_i^n = 1$.

Finally, the distributions obtained from the MCMC scan will be compared to the corresponding prior distributions, which include the ensemble of parameter points without applying the above constraints, i.e.\ stemming from a pure random scan. 

%% file: results.tex
% =====================================
\section{Results and discussion}
\label{Sec:Results}
% =====================================

% --------------------------------------------------------------
\subsection{Couplings between leptons and additional fields}
\label{Subsec:ResultsCouplings}
% --------------------------------------------------------------

Let us recall that the couplings $g_{\Psi}^{\alpha}$ and $g_{\Sigma_{1,2}}^{\alpha}$ ($\alpha = e, \mu, \tau$) introduced in Eq.\ \eqref{Eq:InteractionLagrangian} are computed through the Casas-Ibarra parametrization as discussed in Sec.\ \ref{Subsec:Interactions}. In contrast, the remaining couplings $g_R^{\alpha}$ ($\alpha=e,\mu,\tau$) are mainly constrained by LFV observables summarised in Table \ref{Tab:constraints}. 

\begin{figure}
    \centering
    \includegraphics[scale=0.49]{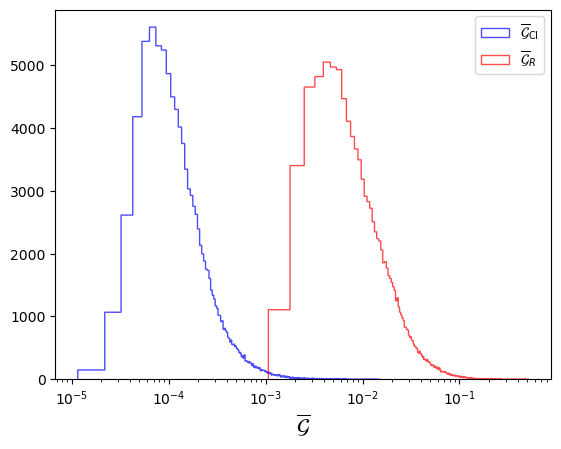}
    \caption{Histograms of the geometrical mean values of couplings entering the Casas-Ibarra parametrisation of the neutrino sector ($\overline{\cal G}_{\text{CI}}$) and couplings not entering the neutrino mass calculation ($\overline{\cal G}_R$) obtained from the MCMC scan.}
    \label{Fig:Couplings}
\end{figure}

In order to understand the respective characteristic order of magnitude for each type of couplings, we display in Fig.\ \ref{Fig:Couplings} the distributions of the respective geometrical mean values,
\begin{align}
    \begin{split}
    \overline{\cal G}_{\text{CI}} ~&=~ \Big( g_{\Psi}^e \, g_{\Psi}^{\mu} \, g_{\Psi}^{\tau} \, g_{\Sigma_1}^e \, g_{\Sigma_1}^{\mu} \, g_{\Sigma_1}^{\tau} \, g_{\Sigma_2}^e \, g_{\Sigma_2}^{\mu} \, g_{\Sigma_2}^{\tau} \Big)^{1/9} \,, \\
    \overline{\cal G}_R ~&=~ \Big( g_R^e \, g_R^{\mu} \, g_R^{\tau} \Big)^{1/3} \,.
    \end{split}
\end{align}
Here, the notation of the couplings implies the absolute value.
As expected, the couplings involved in the neutrino mass generation turn out to be rather small, their mean value peaking around $\overline{\cal G}_{\text{CI}} \sim 10^{-4}$. This relative smallness is required to satisfy the neutrino mass constraints. In contrast, the couplings which are only constrained by LFV processes have more freedom and are generally about two orders of magnitude larger, their mean peaking around $\overline{\cal G}_R \sim 10^{-2}$. LFV processes will therefore be dominated by the couplings $g_R^{\alpha}$ not entering the neutrino mass calculation. 

\begin{figure}
    \centering
    \includegraphics[scale=0.49]{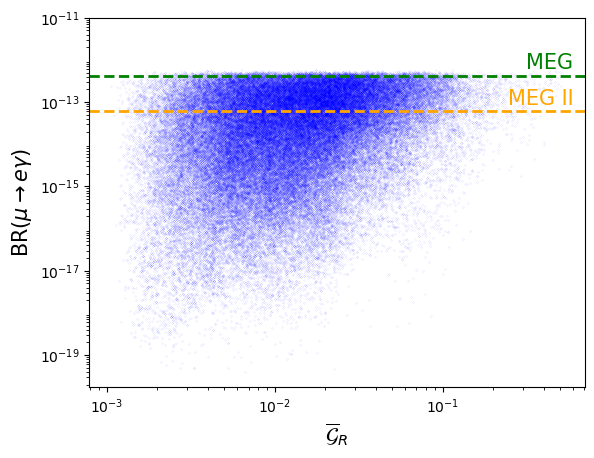}
    \includegraphics[scale=0.49]{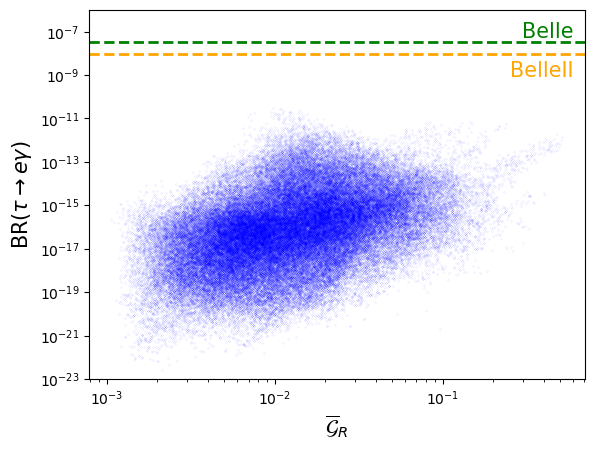}
    \caption{Left: distribution of the absolute values of the coupling parameters $g_R^i$ with ($i = e, \mu$) versus BR($\mu \to e \gamma$). Right: the other generation of $g_R^i$ couplings with ($i = e, \tau$) versus BR($\tau \to e \gamma$). Both graphics are obtained from the Markov Chain Monte Carlo analysis.}
    \label{Fig:Scatter_BRMEG_BRT3E_gR}
\end{figure}

As $\mu-e$ transitions are more severely constrained as compared to $\tau-\mu$ and $\tau-e$ transitions, the couplings the couplings to electrons and muons are more constrained. Phenomenologically viable parameter points typically feature $g_R^{e,\mu} \lesssim 10^{-2}$, but $g_R^{\tau} \lesssim 10^{-1}$. The same effect is exposed in Fig.\ \ref{Fig:Scatter_BRMEG_BRT3E_gR}, where we display the geometrical mean value $\overline{\cal G}_R$ of viable parameter points obtained from the MCMC scan against the respective branching ratios of $\mu\to e\gamma$ and $\tau\to e\gamma$. While part of the parameter space is excluded by the current $\mu\to e\gamma$ constraint, the obtained points are about at least four orders of magnitude below the current $\tau\to e\gamma$ limit. It is to be noted that a large part of the obtained parameter space will be accessible to future searches for the $\mu\to e\gamma$ decay. 

% --------------------------------------------------------------
\subsection{Dark matter mass and phenomenology}
\label{Subsec:ResultsDM}
% --------------------------------------------------------------

The present model features three possible DM candidates, namely the lightest neutral fermion $\chi^0_1$, the lightest neutral scalar $\phi^0_1$, and the pseudoscalar $A^0$. Due to the mixing discussed in Sec.\ \ref{Subsec:Scalars}, the lightest neutral scalar can be either dominated by its singlet component or its doublet component. The same is true for the lightest neutral fermion, which can be either doublet dominated or triplet dominated, see Sec.\ \ref{Subsec:Fermions} concerning the fermion mixing. In the following, we will focus on the phenomenologically viable DM mass ranges in view of the different applied constraints. 

\begin{figure}
    \centering
    \includegraphics[width=0.49\textwidth]{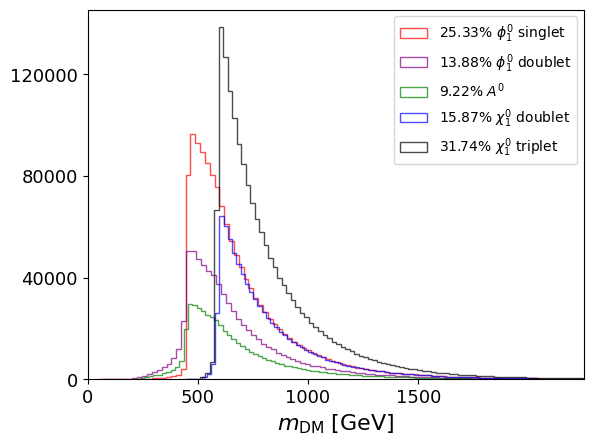}~~~~
    \includegraphics[width=0.471\textwidth]{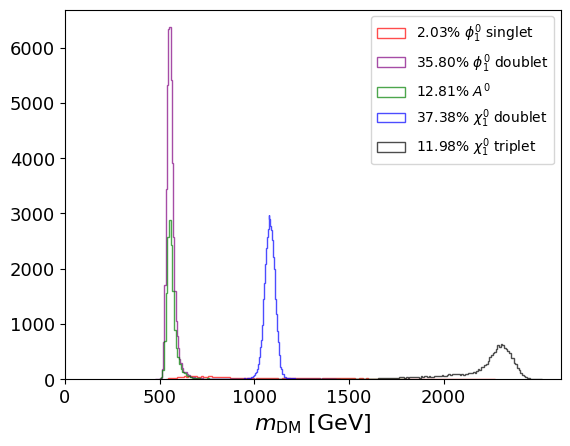}
    \caption{Histograms of the DM masses obtained from a random scan (left) and from our MCMC analysis (right). The different colours indicate the DM nature in terms of scalar, pseudoscalar or fermion as well as singlet, doublet, or triplet domination.}
    \label{Fig:MassesDM}
\end{figure}

In Fig.\ \ref{Fig:MassesDM}, we display the DM mass distributions obtained from a pure random scan, i.e.\ without applying the constraints discussed in Sec.\ \ref{Sec:Constraints}, and from our Markov Chain Monte Carlo scan, i.e.\ including all the constraints. In each case, we indicate explicitly the nature of the respective DM candidate in terms of scalar singlet, scalar doublet, pseudoscalar, fermionic doublet or fermionic triplet dominance. From the pure random scan, i.e.\ without imposing the constraints on relic density and LFV, the DM mass would be expected around 500 GeV to 1 TeV for most of the parameter points. This is explained by the fact that, without imposing the constraints on relic density and LFV, the DM particle is naturally found towards the lower end of the scanned mass intervals (see Table \ref{Tab:setup}).

\begin{figure}
    \centering
    \includegraphics[width=0.49\textwidth]{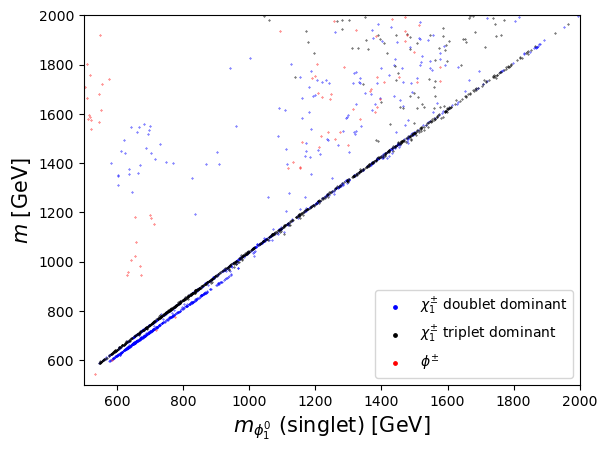}
    \includegraphics[width=0.49\textwidth]{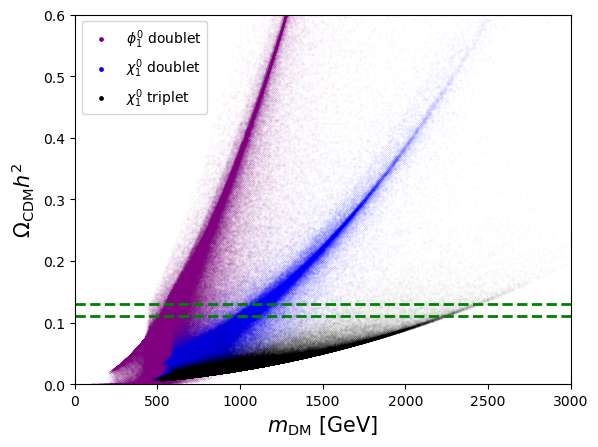}
    \caption{Left: Correlation between the masses of the singlet scalar and the lightest charged particle masses ($\phi^{\pm}$ or $\chi_1^{\pm}$, the latter split depending on its exact nature) for the parameter points obtained from our MCMC analysis where the DM candidate is singlet-scalar dominated. Right: Correlation between DM mass and relic density for a set of random parameter points. The colour code indicates the respective DM nature (scalar singlet not displayed as not relevant for the related discussion, pseudoscalar behaves in the same way as doublet-dominated scalar).}
    \label{Fig:relic_vs_MX1_Random}
\end{figure}

The situation changes drastically when imposing the constraints, in particular the condition of meeting the DM relic density, see Eq.\ \eqref{Eq:RelicDensity}. Depending on the nature of the DM candidate (in terms of spin and composition), its mass is expected within quite precise intervals. Starting with scalar DM, we first notice that singlet-dominated scalar DM occurs in only about 2\% of the phenomenologically viable parameter points, and without preferred mass range. However, it is interesting to note that for the large majority of such configurations, the required DM relic density is achieved through co-annihilations, mainly with either the fermionic doublet or triplet. This is illustrated in the left panel of Fig.\ \ref{Fig:relic_vs_MX1_Random}, where we show the singlet scalar DM mass against the mass of the lightest charged particle, the latter representing de facto either the scalar doublet mass, fermionic doublet mass, or fermionic triplet mass. As can be seen, for most of the points, the fermionic doublet or triplet is very close in mass to the scalar singlet, leading to dominant co-annihilations. As such a configuration is less ``natural'' than, e.g., co-annihilations within a given doublet or triplet (see below), the percentage of parameter points featuring singlet scalar DM is relatively low.

Doublet-dominated scalar DM accounts for about 36\% of the viable parameter points, while pseudoscalar DM accounts for about 13\%. Interestingly, the corresponding distributions are rather peaked around 550 GeV. This is explained by the fact that, stemming from the scalar doublet, the two DM candidates $\phi^0_1$ and $A^0$ are very close in mass. In addition, the charged scalars $\phi^{\pm}$, also stemming from the doublet, have a very similar mass as well. These parameter configurations are dominated by co-annihilations between the three states stemming from the doublet. Consequently, such processes being dominated by gauge couplings, the annihilation cross-section and thus the relic density depends essentially on the mass of the doublet (higher masses lead to smaller annihilation cross-sections and thus to higher values of the relic density). This is illustrated in right panel of Fig.\ \ref{Fig:relic_vs_MX1_Random}, where we display the parameter points, obtained from a random scan in the plane of the DM mass and the relic density. The scalar doublet points are indicated in purple colour. The mass range where the constraint of Eq.\ \eqref{Eq:RelicDensity} is met, is situated in the range around 550 GeV.

Similar features are observed for fermionic DM. While in the pure random scan, the prominent mass range would be around 700 to 900 GeV, two distinct peaks are observed when imposing the constraints. The first peak, situated around $1\,080$ GeV corresponds to doublet-dominated fermionic DM, while the second peak around $2\,300$ GeV corresponds to triplet-dominated DM. Both peaks correspond to co-annihilation regions, in the same way as explained for the scalar doublet above. In this case, the mass dependence of the cross-section is less pronounced, leading to higher mass ranges as compared to the scalar case. Moreover, the triplet containing more states than the doublet, the co-annihilation cross-section is increased in that case, needing a higher DM mass to obtain the same relic density of $\Omega_{\chi}h^2 \sim 0.12$. Again, this is visualised in right panel of Fig.\ \ref{Fig:relic_vs_MX1_Random}, where the fermionic DM points are indicated by blue (doublet) and black (triplet) colour. Let us finally notice that the doublet-dominated peak is exactly the same as the one observed and discussed in Ref.\ \cite{Sarazin:2021nwo}.

It should be noted that the Sommerfeld effect has not been included in the present analysis. For scenarios where the masses of charged particles are close to the DM mass, the Sommerfeld effect could become relevant. However, as shown in Ref.\ \cite{Beneke:2020vff}, the resulting modification to the relic density is not expected to alter the final conclusions. Incorporating Sommerfeld factors into the computation of the relic density may lead to a shift in the peak observed in Fig.\ \ref{Fig:MassesDM}, but this would not impact the overall conclusions of this work.

\begin{figure}
    \centering
    \includegraphics[width=0.75\textwidth]{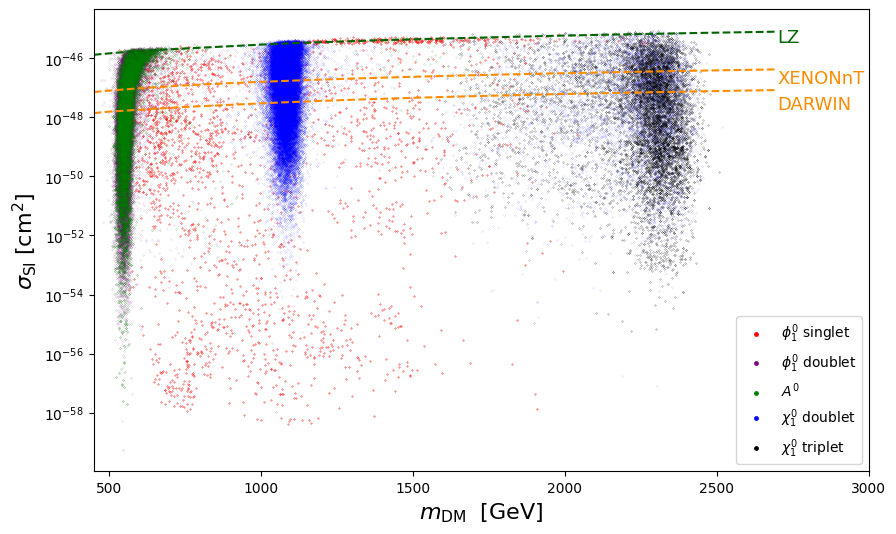}
    \caption{Projection of the spin-independent cross section on the DM mass. The colour code indicates the respective nature of the DM candidate. We indicate the current LZ \cite{LZ:2022lsv} limit (green-dashed line) as well as the expected future limits of the DARWIN \cite{DARWIN:2016hyl} and XENONnT \cite{XENONnT2020} experiments (orange-dashed lines).}
    \label{Fig:DirectDetection}
\end{figure}

Finally, Fig.\ \ref{Fig:DirectDetection} provides information on the spin-independent cross section used for DM direct detection. Note that four main peaks are observed, driving the cross section to lower values. Those peaks are similar to the ones observed Fig. \ref{Fig:MassesDM}, and correspond to the same mass range. The orange-dashed curves are the projections of the future limits of XENONnT \cite{XENONnT2020} and DARWIN \cite{DARWIN:2016hyl} experiments. The values from LZ experiment \cite{LZ:2022lsv} are used for the MCMC and is represented by the green-dashed curve. A large fraction of the points are not discarded by the future experiments. Moreover, the future limits do not give much information on the DM nature and the phenomenology of the model stays large.

Note that our analysis does not include constraints stemming from indirect detection. We have, however, verified that the corresponding cross-sections remain below the corresponding limits given in Ref.\ \cite{Fermi-LAT:2016uux}. In conclusion, the indirect detection of dark matter does not affect the Likelihood and the results in this specific study. Note also that the realistic numerical implementation of indirect detection constraints in a framework involving several additional fields is quite heavy as concluded in Ref.\ \cite{Armand:2022sjf}.

% --------------------------------------------------------------
\subsection{Lepton flavour violating transitions}
\label{Subsec:ResultsLFV}
% --------------------------------------------------------------

\begin{figure}
    \centering
    \includegraphics[width=0.49\textwidth]{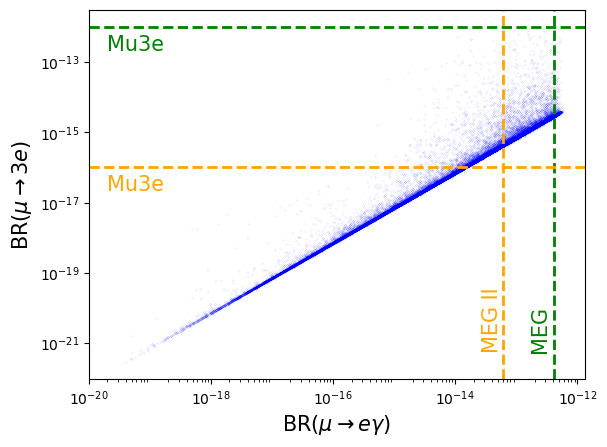}
    \includegraphics[width=0.48\textwidth]{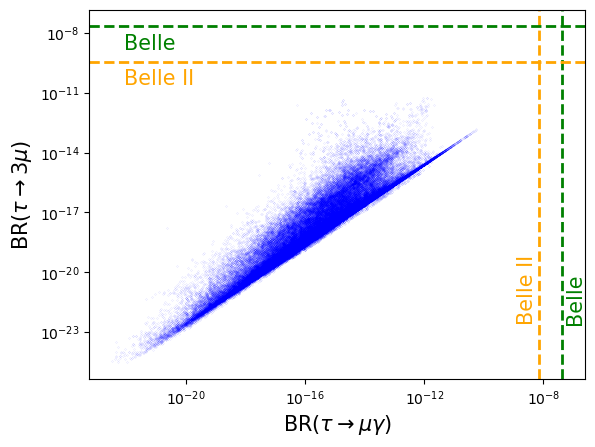}
    \caption{Left: Correlation of the respective branching ratios of the decays $\mu\to e\gamma$ and $\mu\to 3e$ obtained from our MCMC analysis. Right: Same for $\tau\to \mu\gamma$ and $\tau\to 3\mu$. In both figures, we indicate current (green-dashed lines) and future (orange-dashed lines) experimental limits.}
    \label{Fig:Scatter_BRMEG_BrM3E}
\end{figure}

As already mentioned above, LFV transitions are present in scotogenic models and, at the same time, important probes allowing to constrain the associated parameter space. Given the respective experimental accuracies, $\mu - e$ transitions are the most constraining observables in this respect. In Fig.\ \ref{Fig:Scatter_BRMEG_BrM3E}, we illustrate our findings for the branching ratios of the decays $\mu\to e\gamma$, $\mu\to 3e$, $\tau\to \mu\gamma$, and $\tau\to 3\mu$, obtained from our MCMC analysis.

Starting with the $\mu-e$ transition, the dominant constraint clearly stems from the MEG measurement of the $\mu \to e \gamma$ decay, constraining the couplings $g_R^e$ and $g_R^{\mu}$, as well as the couplings $y$ appearing in Eq.\ \eqref{Eq:FermionLagrangian}. As already discussed above, the couplings $g_{\psi,\Sigma_{1,2}}^{e,\mu,\tau}$ involved in the neutrino mass generation are generally numerically smaller. It is to be noted that future searches for $\mu-e$ transitions will be able to challenge a relatively large number of currently viable parameter points. 

For the $\mu-e$ transition, there is an approximate linear correlation between the two branching ratios, which, however, becomes washed out at larger values. This can be traced to the fact that certain contributions to $\mu \to 3e$ are related to the decay $\mu \to e\gamma$, as the photon can decay into a pair $e^+e^-$. Additional contributions to $\mu\to 3e$ stem from box diagrams, which are less important in case of smaller coupling values. For increasing couplings, the branching ratios increase, and the relative box contribution to $\mu\to 3e$ increases as well, leading to the wash-out of the correlation. The correlation is less pronounced in the case of the $\tau - \mu$ and $\tau - e$ transitions, as the couplings to tau leptons are much less constrained, and consequently box contributions are generally more important. The example of $\tau \to \mu\gamma$ and $\tau\to 3\mu$ is displayed in Fig.\ \ref{Fig:Scatter_BRMEG_BrM3E}.

\begin{figure}
    \centering
    \includegraphics[scale=0.45]{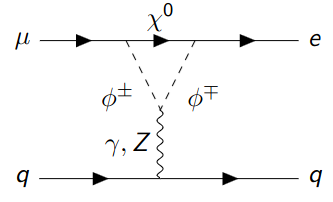}
    \caption{Example of penguin diagram of the $\mu-e$ conversion in a scotogenic framework.}
    \label{Fig:mueconversionPenguin}
\end{figure}

\begin{figure}
    \centering
    \includegraphics[width=0.49\textwidth]{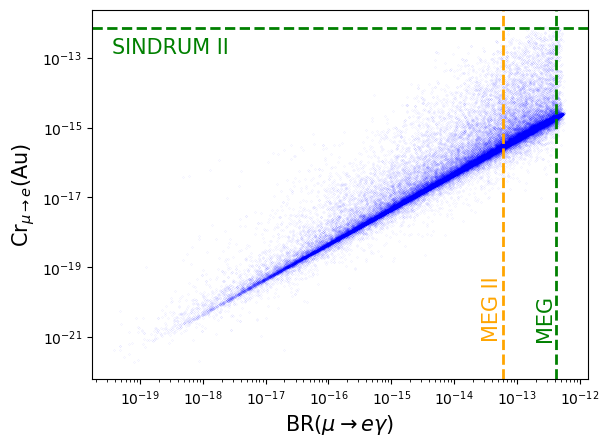}
    \includegraphics[width=0.48\textwidth]{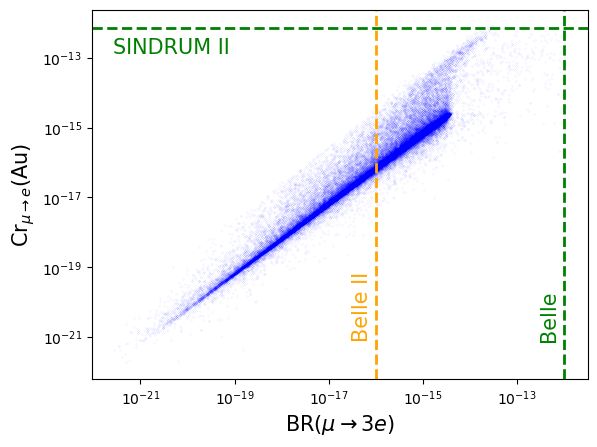}
    \caption{Correlation between the $\mu-e$ conversion rate in Au nuclei and branching ratios of $\mu \to e \gamma$ (left) and $\mu \to 3e$ (right) obtained from the MCMC analysis.}
    \label{Fig:MEG_CRAU}
\end{figure}

In addition to rare muon decays, the couplings related to $\mu-e$ transitions can be constrained by data on such transitions within atomic nuclei. For a given atom, the associated conversion rate depends on the effective atomic charge and the atomic number of the considered atom, the total muon capture rate, and form factors generated by dipole photon penguins as shown in Fig.\ \ref{Fig:mueconversionPenguin} \cite{Boruah:2021qlf}. The loop in the diagram includes the couplings $g_{\Psi}^{e,\mu}$, $g_{\Sigma_{1,2}}^{e,\mu}$, and $g_R^{e,\mu}$. In Fig.\ \ref{Fig:MEG_CRAU}, we display the $\mu-e$ conversion rate with gold atoms in relation to the $\mu\to e\gamma$ and $\mu\to 3e$ branching ratios. As can be seen, current data cannot compete with the precision of the rare decay searches. However, it is to be noted that future searches, e.g.\ COMET experiment, for $\mu-e$ conversion within Aluminium atoms will greatly improve sensitivity to about $10^{-15} - 10^{-17}$ \cite{COMET2024}. Let us finally mention that the correlation is again washed out for higher coupling values. Consequently, the constraints imposed by their relationship become more complex and require careful consideration. This suggests that in such regimes, the interplay between the two observables cannot be neglected and must be analysed jointly to understand their combined impact accurately.

% -------------------------------------
\subsection{Comment on $(g-2)_{\mu}$ and EDM}
\label{Subsec:gminus2}
% -------------------------------------

A further observable in the leptonic sector, although not involving flavour violating processes, is the anomalous magnetic moment, quantifying the deviation of a lepton's magnetic moment from the value expected from the Dirac equation based on charge and spin. Recently, tensions have been reported between the Standard Model prediction of the anomalous magnetic moment of the muon and its experimental measurement \cite{g2Exp}, constraining new physics contributions to the interval \cite{g-2SM}
\begin{align}
    a_{\mu}^{\text{BSM}} ~=~ \frac{\big( g-2 \big)_{\mu}^{\text{BSM}}}{2}~=~ \big( 251 \pm 59 \big) \times 10^{-11} \,.
\end{align}
Although we do not include this observable as a constraint in our Markov Chain Monte Carlo analysis, it is interesting to investigate to which extent the model under consideration may explain the above tension. For the obtained viable parameter points, the contribution stemming from the additional fields is generally too low to explain the observed tension. This can be traced to the relative smallness of the involved couplings $g_{\Psi}^{\mu}$, $g_{\Sigma}^{\mu}$, and $g_R^{\mu}$.

Let us note that it is possible to obtain larger contributions within the model under consideration. This would require a specific coupling hierarchy, which may be achieved by introducing a special hierarchy in the angles of the rotation matrix $R$ appearing in the Casas-Ibarra parametrization given in Eq.\ \eqref{Eq:CouplingMatrixCasasIbarra}, according to the method developed in Ref.\ \cite{Alvarez:2023dzz}. The implementation of this method is, however, beyond the scope of the present study.

A related observable to $(g-2)$ is the electric dipole moment (EDM). While the Standard Model contributions for the lepton EDMs are CKM-suppressed, new physics contributions are constrained by the upper limits given in Table \ref{Tab:constraints}. In the present setup, as already mentioned above, the smallness of the involved couplings leads to a suppression of additional contributions to the lepton EDMs. The impact of these constraints related to $CP$-violation remain therefore negligible.

% --------------------------------------------------------------
\subsection{Mass distributions and collider signatures}
\label{Subsec:ResultsMasses}
% --------------------------------------------------------------

\begin{figure}
    \centering
    \includegraphics[width=0.5\textwidth]{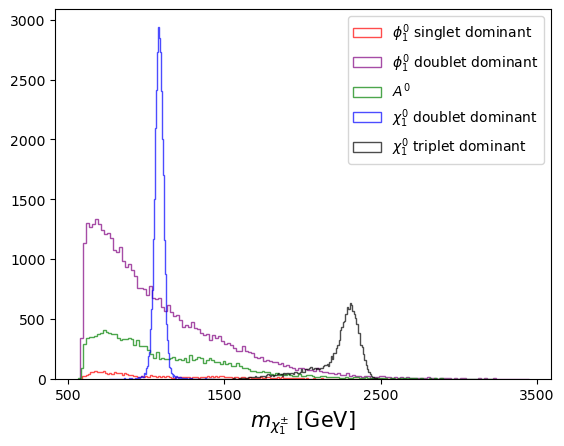}
    \includegraphics[width=0.49\textwidth]{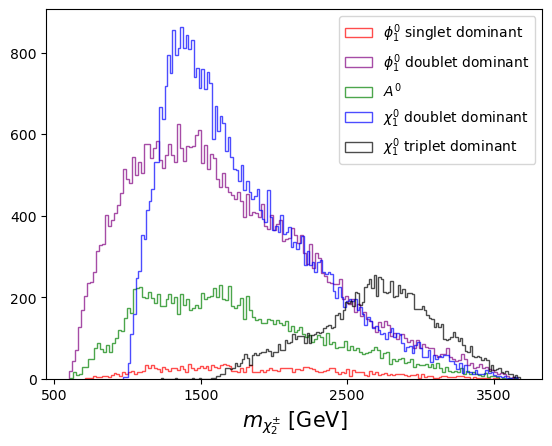}
    \caption{Mass distribution of the two lightest charged fermions ($\chi^{\pm}_1$ and $\chi^{\pm}_2$) obtained from our analysis. The colour code indicates the nature of the corresponding DM candidate.}
    \label{Fig:ChargedX12}
\end{figure}

Finally, we discuss the mass distributions obtained from our MCMC analysis in view of potential collider signatures. Let us recall that, as shown in Fig.\ \ref{Fig:MassesDM} and discussed in Sec.\ \ref{Subsec:DM}, we obtain distinct mass regions for the DM candidate, depending on its precise nature and dominant component. In Fig.\ \ref{Fig:ChargedX12}, we present the corresponding mass distributions of the charged fermions, $m_{\chi_1^{\pm}}$ and $m_{\chi_2^{\pm}}$, obtained from the MCMC analysis, and again separated depending on the respective DM candidate. 

In case of scalar DM, the mass distribution of the charged fermions closely resembles that of the neutral particles presented in Fig.\ \ref{Fig:MassesDM} for a random analysis without constraints. This indicates that the masses of the charged fermions are largely unaffected by the scalar nature of the DM. Conversely, when the DM candidate is fermion-like, the mass distribution of the lightest charged fermion exhibits two distinct peaks located at about 1080 GeV and 2300 GeV, corresponding to the peaks observed for the mass of the lightest neutral fermion in Fig.\ \ref{Fig:MassesDM}. As for both doublet and triplet DM, the charged fermions feature a mass close to the DM candidate, the same peaks are observed for the lightest charged fermion. A similar mass distribution is observed for $\chi_2^{\pm}$, albeit with a broader shape, highlighting the sensitivity of the charged fermion masses to the nature of the DM candidate. 

\begin{figure}
    \centering
    \includegraphics[width=0.49\textwidth]{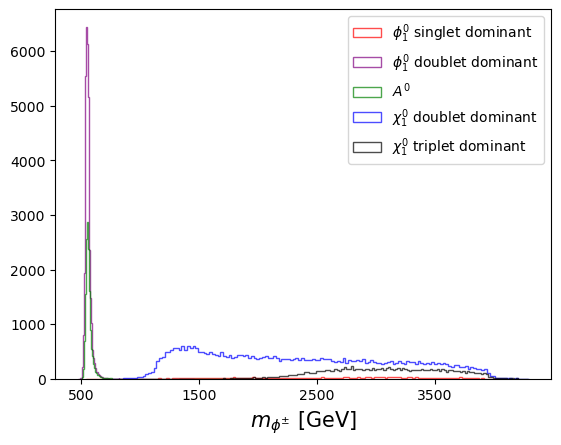}
    \caption{Mass distribution of the charged scalar ($\phi^{\pm}$) obtained from our MCMC analysis. The colour code indicates the nature of the corresponding DM candidate.}
    \label{Fig:ChargedX23}
\end{figure}

Coming to the mass distributions of the charged scalar shown in Fig.\ \ref{Fig:ChargedX23}, a clear peak is observed at around 550 GeV for the case of doublet-like DM ($\phi^0_1$ or $A^0$). Again, this corresponds to the peak of Fig.\ \ref{Fig:MassesDM}, as in this case the DM and the charged scalar both stem from the doublet. For the remaining cases, the mass distribution of the charged scalar does not exhibit particular features and favours values between 1 and 4 TeV.

It is interesting to note that, independently of the exact DM nature, we can expect charged particles at essentially the same mass as the DM particle. As explained above, this is due to the importance of co-annihilations, which are, in the present model, the only available mechanism to meet the DM relic density constraint of Eq.\ \eqref{Eq:RelicDensity}.

Concerning collider signatures, this configuration has two interesting aspects. First, charged scalars or fermions featuring a numerically more important production cross-section, information about the mass of charged states may give a hint towards the DM mass. Moreover, the rather small mass gap between the DM candidate and the charged particle, or between the charged particle and the associated neutral states from the same doublet or triplet, may lead to signatures of so-called long-lived particles, the charged state featuring a more important production cross-section but, due to the small mass gap, a relatively long life time. Due to the rather predictive phenomenology of the model, dedicated studies of such signatures will be interesting to carry out. They are, however, beyond the scope of the present paper.

% --------------------------------------------------------------
\subsection{Reference scenarios}
\label{Subsec:ExampleScenarios}
% --------------------------------------------------------------

In this section, we present five selected reference scenarios, which represent the different DM configurations discussed in the previous sections. The given parameter configurations correspond to the highest likelihoods obtained for each DM configuration from our MCMC analysis discussed above. For each example, we discuss the mass spectrum and the dominating DM (co-)annihilation channels. The latter are summarized in Table \ref{Tab:Scenarios}.

The selected reference scenarios are selected according to the respective DM phenomenology:
\begin{center}
\begin{tabular}{rcl}
     \bf I &  & Singlet scalar DM ($m_{\text{DM}} = m_{\phi^0_1} \approx M_S$) \\
     \bf II &   & $CP$-even doublet scalar DM ($m_{\text{DM}} = m_{\phi^0_1} \approx M_{\eta}$) \\
     \bf III &   & $CP$-odd doublet scalar DM ($m_{\text{DM}} = m_{A^0} \approx M_{\eta}$) \\
     \bf IV &   & Doublet fermion DM ($m_{\text{DM}} = m_{\chi^0_1} \approx M_{\Psi}$) \\
     \bf V &   & Triplet fermion DM ($m_{\text{DM}} = m_{\chi^0_1} \approx M_{\Sigma_1}$) \\
\end{tabular}
\end{center}
The associated data can be found in SLHA format \cite{SLHA1, SLHA2} as ancillary files with the electronic submission of the present paper. The given scenarios may be used as typical phenomenologically viable parameter configuration, e.g.\ in future collider or other studies.

\begin{table}[]
    \centering
    \begin{tabular}{|c||c|c|c|c|c|}
        \hline
           & \bf \!\!\! Scenario I \!\!\!& \!\!\! \bf Scenario II \!\!\! & \!\!\! \bf Scenario III \!\!\! & \!\!\! \bf Scenario IV \!\!\! & \!\!\! \bf Scenario V \!\!\! \\
           & \footnotesize(sing.\ scal.) & \footnotesize(doubl.\ scal.) & \footnotesize(pseudo-scal.) & \footnotesize(doubl.\ ferm.) & \footnotesize(tripl.\ ferm.) \\
        \hline
        \hline
        $m_{\phi^0_1}$ [GeV] & 738.00 & 560.08 & 553.55 & 1370.37 & 2818.68 \\
        $m_{A^0}$ [GeV] & 3698.02 & 561.13 & 553.12 & 3693.55 & 2820.25 \\
        $m_{\phi^{\pm}_1}$ [GeV] & 3698.19 & 561.01 & 553.68 & 3693.75 & 2820.57 \\
        \hline
        $m_{\chi^0_1}$ [GeV] & 748.72 & 955.62 & 972.36 & 1089.90 & 2303.39 \\
        $m_{\chi^{\pm}_1}$ [GeV] & 751.88 & 955.72 & 972.55 & 1090.19 & 2303.41 \\        
        \hline
        \hline
        $m_{\text{DM}}$ [GeV] & 738.00 & 560.08 & 553.12 & 1089.90 & 2303.39 \\
        $\Delta m$ [GeV] & 13.88 & 0.93 & 0.56 & 0.29 & 0.02 \\
        \hline
        $\sigma_{\text{SI}}$ [cm$^2$] & $6.6\times 10^{-49}$ & $8.8\times 10^{-48}$ & $9.1\times 10^{-49}$ & $3.6\times 10^{-46}$ & $1.3\times 10^{-51}$ \\
        \hline
        \hline
        BR($\mu \to e\gamma$) & $5.2\times 10^{-15}$ & $4.9\times 10^{-16}$ & $2.6\times 10^{-16}$ & $3.5\times 10^{-14}$ & $1.3\times 10^{-16}$ \\
        \hline
        \hline
        $\phi^0_1 \phi^0_1 \to VV$ & -- & 27\% & 16\% & -- & -- \\    
        $A^0 A^0 \to VV$ & -- & 14\% & 23\% & -- & -- \\    
        $\phi^0_1 \phi^{\pm} \to VV$ & -- & 12\% & 11\% & -- & -- \\
        $A^0 \phi^{\pm} \to VV$ & -- & 10\% & 12\% & -- & -- \\
        $\phi^{+} \phi^{-} \to VV$ & -- & 29\% & 29\% & -- & -- \\     
        \hline
        $\chi^0_1 \chi^0_1 \to VV$ & 3\% & -- & -- & 2\% & 8\% \\
        $\chi^0_1 \chi^0_2 \to q\bar{q}$ & 15\% & -- & -- & 12\% & -- \\
        $\chi^0_1 \chi^0_2 \to \nu \bar{\nu}$ & 3\% & -- & -- & 3\% & -- \\
        $\chi^0_1 \chi^{\pm}_1 \to q \bar{q}$ & 24\% & -- & -- & 21\% & 35\% \\ 
        $\chi^0_1 \chi^{\pm}_1 \to \ell \nu$ & 8\% & -- & -- & 7\% & 9\% \\ 
        $\chi^0_2 \chi^0_2 \to VV$ & 1\% & -- & -- & 2\% & -- \\
        $\chi^0_2 \chi^{\pm}_1 \to q \bar{q}$ & 21\% & -- & -- & 21\% & -- \\
        $\chi^0_2 \chi^{\pm}_1 \to \ell \nu$ & 3\% & -- & -- & 7\% & -- \\
        $\chi^{\pm} \chi^{\pm} \to q \bar{q}$ & 9\% & -- & -- & 12\% & 18\% \\
        $\chi^{\pm} \chi^{\pm} \to \ell \bar{\ell}$ & 3\% & -- & -- & 3\% & -- \\
        $\chi^{\pm} \chi^{\pm} \to VV$ & --  & -- & -- & 16\% & 22\% \\
        \hline
        \end{tabular}
    \caption{Masses of lightest neutral and charged scalars and fermions, DM mass ($m_{\text{DM}}$), mass difference between the DM candidate and the lightest charged particle ($\Delta m = m_{\pm}-m_{\text{DM}}$), branching ratio of the $\mu\to e\gamma$ decay, spin-independent DM direct detection cross-section ($\sigma_{\text{SI}}$), and relative contributions of the dominating DM (co-)annihilation channels for the five reference scenarios based on our MCMC analysis. The channels are regrouped into classes according to $V\in\{\gamma, Z^0, W^{\pm}\}$, $q \in \{ u,d,c,s,t,b \}$, $\ell \in \{ e,\mu,\tau \}$, and $\nu \in \{ \nu_e,\nu_{\mu},\nu_{\tau} \}$. Contributions less than one percent are not displayed. The five scenarios meet the relic density and Higgs mass constraints.}
   \label{Tab:Scenarios}
\end{table}

Concerning the annihilation channels, as already stated above, co-annihilations are the dominant mechanism to achieve the required relic density. We also note that co-annihilations contribute in many cases more strongly to the total annihilation cross-section than pair annihilation of the DM particle.

Coming to possible collider signatures, the possibility of long-lived particles mentioned above seems to be realistic for Scenarios II, III, IV, and V, while the mass gap for the first remaining scenario is too important. Note, however, that the production cross-section of the charged fermions is rather low for Scenarios IV and V due to the large fermion mass. 

%% file: conclusion.tex
%=====================================
\section{Conclusion}
\label{Sec:Conclusion}
%=====================================

The so-called "T1-2G" model is a rather general scotogenic framework, including singlet, doublet and triplet fields, naturally generating three non-zero neutrino masses, providing three viable dark matter  candidates, and offering a wide range of phenomenological features. In this work, we have provided a comprehensive analysis of the associated parameter space, incorporating constraints from the Higgs sector, neutrino masses, lepton-flavour violating processes, and dark matter . To the best of our knowledge, this is the first thorough investigation of the "T1-2G" model considering the entire parameter space while accounting for such a diverse set of constraints. Our analysis is mainly based on a Markov Chain Monte Carlo numerical scan, while additional information is obtained from a pure random scan without imposing constraints.

We make use of the Casas-Ibarra parametrization to compute the couplings between left-handed leptons, new scalars, and new fermions, taking into account current neutrino data \cite{NuFit2020}. The remaining coupling parameters, essentially those involving right-handed leptons, are severely constrained by lepton-flavour violating processes, such as the $\mu\to e\gamma$ decay and $\mu-e$ conversion in atomic nuclei. 

Further strong constraints stem from the requirement that the model should explain the cold dark matter  present in the Universe. Throughout the viable parameter space, and independently of the nature of the dark matter  candidate, dark matter  annihilation is driven by important co-annihilations. This feature yields very specific mass ranges for the dark matter  particle. For scalar dark matter , a very large fraction of viable points is situated around a dark matter  mass of 550 GeV, while the mass peaks for fermionic dark matter  are around 1080 GeV and 2300 GeV. Additionally, charged particles are expected in the same mass ranges, making the model very predictive concerning the mass spectrum of the new states, and the associated collider signatures. The latter may include signatures similar to electroweakino production in supersymmetric models or, due to small mass differences in co-annihilation scenarios, signatures related to long-lived particles.

Concerning dark matter  direct searches, while the model allows to evade the cross-section limits stemming from recent searches, upcoming experimental improvements, e.g.\ the XENON-nT experiment \cite{XENONnT2020}, will be able to challenge a large part of the viable parameter space. However, certain regions of viable parameter points featuring very low scattering cross-sections will remain immune to direct searches. A similar comment can be made concerning lepton-flavour violating transitions. The current constraints, mainly on the $\mu-e$ transition, will be improved in a near future thanks to the MEG II experiment \cite{Meucci:2022qbh} and the future search for $\mu-e$ conversion in gold atoms \cite{COMET2024}. Although the associated data may challenge an important part of the now allowed parameter space, a large fraction of parameter configurations features transition rates which are several orders of magnitude below the projected limits.

Let us note that the model may explain the deviation of the anomalous magnetic moment of the muon from the Standard Model prediction, as it includes sufficient degrees of freedom to meet the latest results. This requires, however, a specific hierarchy between the couplings to electrons and to muons \cite{Alvarez:2023dzz}. The model under consideration may also provide an explanation to the baryon-antibaryon asymmetry observed in the Universe via leptogenesis \cite{Alvarez:2023dzz}. A detailed analysis of the latter points is, however, left for future work.

Lastly, our results may serve as a useful guide for future collider studies. To this end, we provide five typical phenomenologically viable parameter configurations representing the different possible configurations for dark matter  in the model. Moreover, the associated mass spectra yield different signatures. For example, experimental searches for electroweakinos in supersymmetric extensions of the Standard Model particles may provide approximate exclusion limits for the fermion sector of the present model. Moreover, numerous parameter configurations will lead to long-lived particles. Such dedicated studies are left for future investigations.

%% file: appendix.tex
%=====================================
\section{Conditions on scalar couplings parameters}
\label{App:ScalarConditions}
%=====================================

To insure vacuum stability, the potential has to be bounded from below. This translates into the following conditions on the couplings
\begin{equation}
    \begin{split}
        2 \lambda_H ~&>~ \frac{M^2_H}{M^2_{\eta}} \left(\lambda_{\eta} + \lambda^{'}_{\eta} + \lambda^{''}_{\eta}\right) \,, \\
        2 \lambda_H M^2_S ~&>~ \lambda_S M^2_H \,, \\
        \lambda_{\eta} ~&>~ -2 \sqrt{\lambda_H \lambda_{4 \eta}} \,, \\
        \lambda_{\eta} + \lambda^{'}_{\eta} - \left|\lambda^{''}_{\eta}\right| ~&>~ -2 \sqrt{\lambda_H \lambda_{\eta}} \,, \\
        \frac{1}{2} \lambda_S ~&>~ -2 \sqrt{\lambda_H \lambda_{4S}} \,, \\
        \frac{1}{2} \lambda_{S\eta} ~&>~ -2 \sqrt{\lambda_{4\eta} \lambda_{4S}} \,.
    \end{split}
    \label{Eq:BoundednessFromBelow1}
\end{equation}
The scalar couplings also have to satisfy
\begin{equation}
    \begin{split}
        \sqrt{8 \lambda_H \lambda_{4\eta} \lambda_{4S}} + \lambda_{\eta} \sqrt{2 \lambda_{4S}} + \frac{1}{2} \lambda_{S} \sqrt{2 \lambda_{4 \eta}} + \frac{1}{2} \lambda_{S \eta} \sqrt{2 \lambda_H} + \rho_1 ~&>~ 0 \,, \\
        \sqrt{8 \lambda_H \lambda_{4\eta} \lambda_{4S}} + \left(\lambda_{\eta} + \lambda^{'}_{\eta} - \big|\lambda^{''}_{\eta}\big|\right) \sqrt{2 \lambda_{4S}} + \frac{1}{2} \lambda_{S} \sqrt{2 \lambda_{4 \eta}} + \frac{1}{2} \lambda_{S \eta} \sqrt{2 \lambda_H} + \rho_2 ~&>~ 0 \,,
    \end{split}
    \label{Eq:BoundednessFromBelow2}
\end{equation}
with $\rho_{1,2}$ being defined as 
\begin{equation}
    \begin{split}
        \rho_1 ~&=~ \sqrt{2 \left(\lambda_{\eta} + 2 \sqrt{\lambda_H \lambda_{4 \eta}}\right)\left(\frac{1}{2}\lambda_{S} + 2 \sqrt{\lambda_H \lambda_{4S}}\right)} \,, \\
        \rho_2 ~&=~ \sqrt{2 \left(\lambda_{\eta} + \lambda^{'}_{\eta} - \left|\lambda^{''}_{\eta}\right| + 2 \sqrt{\lambda_H \lambda_{4 \eta}}\right)\left(\frac{1}{2}\lambda_{S} + 2 \sqrt{\lambda_H \lambda_{4S}}\right)} \,.
    \end{split}
    \label{Eq:CoefficientsRho}
\end{equation}

%=================================================
\section{Expression of the loop matrix $M_L$}
\label{App:LoopMatrix}
%=================================================

The elements of the $3 \times 3$ matrix $M_L$ appearing in Eq.\ \eqref{Eq:NeutrinoMassMatrix}, expressed in terms of the scalar and fermionic masses and mixing matrices, read
\begin{equation}
    \begin{split}
    \left( M_L \right)_{11} &= \sum_{k,n} b_{kn}  \, \big(U_{\chi^0}^{\dagger}\big)^2_{4k} \, \big(U_{\phi}^{\dagger}\big)^2_{1n} \,, \\
    \left( M_L \right)_{22} &= \frac{1}{2} \sum_{k,n} b_{kn} \, \big(U_{\chi^0}^{\dagger}\big)^2_{1k} \, \Big[ \big(U_{\phi}^{\dagger}\big)^2_{2n} - \big( U_{\phi}^{\dagger} \big)^2_{3n} \Big]  \,, \\
    \left( M_L \right)_{33} &= \frac{1}{2} \sum_{k,n} b_{kn}  \, \big(U_{\chi^0}^{\dagger}\big)^2_{2k} \, \Big[ \big(U_{\phi}^{\dagger}\big)^2_{2n} - \big( U_{\phi}^{\dagger} \big)^2_{3n} \Big] \,, \\
    \left( M_L \right)_{12} = \left( M_L \right)_{21} &= \frac{1}{\sqrt{2}}\sum_{k,n} b_{kn}  \, \big(U_{\chi^0}^{\dagger}\big)_{1k} \, \big(U_{\chi^0}^{\dagger}\big)_{4k} \, \big(U_{\phi}^{\dagger}\big)_{1n} \, \big(U_{\phi}^{\dagger}\big)_{2n} \,, \\
    \left( M_L \right)_{13} = \left( M_L \right)_{31} &= \frac{1}{\sqrt{2}} \sum_{k,n} b_{kn} \, \big(U_{\chi^0}^{\dagger}\big)_{2k} \, \big(U_{\chi^0}^{\dagger}\big)_{4k} \, \big(U_{\phi}^{\dagger}\big)_{1n} \, \big(U_{\phi}^{\dagger}\big)_{2n}  \,, \\
    \left( M_L \right)_{23} = \left( M_L \right)_{32} &= \frac{1}{2} \sum_{k,n} b_{kn}  \, \big(U_{\chi^0}^{\dagger}\big)_{1k} \big(U_{\chi^0}^{\dagger}\big)_{2k} \, \Big[ \big(U_{\phi}^{\dagger}\big)^2_{2n} - \big( U_{\phi}^{\dagger} \big)^2_{3n} \Big] \,, \\
    \end{split}
    \label{Eq:LoopComponents}
\end{equation}
where the sums run over the neutral fermion mass eigenstates $\left(k= 1,2,3,4\right)$ and the neutral scalar mass eigenstates $\left(n = 1,2,3, \; \text{the last one corresponding to the pseudo-scalar $A^0$} \right) $. The coefficients $b_{kn}$, stemming from the loop integrals, are given by 
\begin{equation}
    b_{kn} = \frac{1}{16 \pi^2} \frac{m_{\chi^0_k}}{m^2_{\phi^0_n} - m^2_{\chi^0_k}} \left[ m^2_{\chi^0_k} \ln\!{\big(m^2_{\chi^0_k}\big)} - m^2_{\phi^0_n} \ln\!{\big(m^2_{\phi^0_n}\big)}\right] \,.
    \label{Eq:LoopCoefficients}
\end{equation}
We note that there is no dependence on the renormalisation scale, as the terms given in Eq.\ \eqref{Eq:LoopComponents} constitute the leading order contribution.

%=====================================
\section{Matrices involved in the Casas-Ibarra parametrization}
\label{App:CasasIbarra}
%=====================================

The PMNS matrix can be written as
\begin{equation}
    U_{\text{PMNS}} ~=~ \!\! \begin{pmatrix} 
		    1 & 0 & 0 \\ 0 & c_{23} & s_{23} \\ 0 & -s_{23} & c_{23} 
		\end{pmatrix} \!\!
		\begin{pmatrix} 
		    c_{13} & 0 & s_{13} e^{-i \delta_{\rm CP}} \\ 
		    0 & 1 & 0 \\ 
		    -s_{13}e^{i \delta_{\rm CP}} & 0 & c_{13} 
		\end{pmatrix} \!\!
		\begin{pmatrix} 
		    c_{12} & s_{12} & 0 \\ -s_{12} & c_{12} & 0 \\ 0 & 0 & 1 
		\end{pmatrix} \!\!
		\begin{pmatrix} 
		    1 & 0 & 0 \\ 
		    0 & e^{i \alpha_{1}} & 0 \\ 
		    0 & 0 & e^{i \alpha_{2}}
        \end{pmatrix} \,,
    \label{Eq:PMNSMatrix}
\end{equation}
where $c_{ij}$ and $s_{ij}$ are the cosine and sine of the three angles $\theta_{12}$, $\theta_{13}$, and $\theta_{23}$, respectively. The $CP$-violating phase $\delta_{\rm CP}$ takes its value in the range from Ref.\ \cite{NuFit2020}, and the Majorana phases $\alpha_{1,2}$ are taken between $0$ and $\pi$. The rotation matrix appearing in Eq.\ \eqref{Eq:CouplingMatrixCasasIbarra} can be parametrized as 
\begin{equation} 
     R ~=~ \left( \begin{array}{ccc}
         \sqrt{1-r_1^2 } & -r_1 & 0  \\
         r_1 & \sqrt{1-r_1^2 } & 0 \\
         0 & 0 & 1
     \end{array} \right) \left( \begin{array}{ccc}
         \sqrt{1-r_2^2 } & 0 & r_2  \\
         0 & 1 & 0 \\
         -r_2 & 0 & \sqrt{1-r_2^2 }
     \end{array} \right) \left( \begin{array}{ccc}
         1 & 0 & 0  \\
         0 & \sqrt{1-r_3^2 } & -r_3 \\
         0 & r_3 & \sqrt{1-r_3^2 }
     \end{array} \right)
     \label{eq:Rmat}
\end{equation}
with the three parameters $r_i$ defined in Sec.\ \ref{Subsec:Setup}.

%% file: main.bbl
\providecommand{\href}[2]{#2}\begingroup\raggedright\begin{thebibliography}{10}

\bibitem{Planck2018}
{\bf Planck} , N.~Aghanim et~al., {\it {Planck 2018 results. VI. Cosmological
  parameters}},  \href{http://xxx.lanl.gov/abs/1807.0620}{{\tt
  arXiv:1807.0620}}.

\bibitem{NuFit2020}
I.~Esteban, M.~C. Gonzalez-Garcia, M.~Maltoni, T.~Schwetz, and A.~Zhou, {\it
  {The fate of hints: updated global analysis of three-flavor neutrino
  oscillations}},  {\em JHEP} {\bf 09} (2020) 178,
  [\href{http://xxx.lanl.gov/abs/2007.1479}{{\tt arXiv:2007.1479}}].

\bibitem{Minkowski:1977sc}
P.~Minkowski, {\it {$\mu \to e\gamma$ at a Rate of One Out of $10^{9}$ Muon
  Decays?}},  {\em Phys. Lett. B} {\bf 67} (1977) 421--428.

\bibitem{Yanagida:1980xy}
T.~Yanagida, {\it {Horizontal Symmetry and Masses of Neutrinos}},  {\em Prog.
  Theor. Phys.} {\bf 64} (1980) 1103.

\bibitem{PhysRevLett.56.561}
R.~N. Mohapatra, {\it Mechanism for understanding small neutrino mass in
  superstring theories},  {\em Phys. Rev. Lett.} {\bf 56} (Feb, 1986) 561--563.

\bibitem{Ma:2006km}
E.~Ma, {\it {Verifiable radiative seesaw mechanism of neutrino mass and dark
  matter}},  {\em Phys. Rev. D} {\bf 73} (2006) 077301,
  [\href{http://xxx.lanl.gov/abs/hep-ph/0601225}{{\tt hep-ph/0601225}}].

\bibitem{Toma:2013zsa}
T.~Toma and A.~Vicente, {\it {Lepton Flavor Violation in the Scotogenic
  Model}},  {\em JHEP} {\bf 01} (2014) 160,
  [\href{http://xxx.lanl.gov/abs/1312.2840}{{\tt arXiv:1312.2840}}].

\bibitem{Vicente:2014wga}
A.~Vicente and C.~E. Yaguna, {\it {Probing the scotogenic model with lepton
  flavor violating processes}},  {\em JHEP} {\bf 02} (2015) 144,
  [\href{http://xxx.lanl.gov/abs/1412.2545}{{\tt arXiv:1412.2545}}].

\bibitem{Fraser:2014yha}
S.~Fraser, E.~Ma, and O.~Popov, {\it {Scotogenic Inverse Seesaw Model of
  Neutrino Mass}},  {\em Phys. Lett. B} {\bf 737} (2014) 280--282,
  [\href{http://xxx.lanl.gov/abs/1408.4785}{{\tt arXiv:1408.4785}}].

\bibitem{Baumholzer:2019twf}
S.~Baumholzer, V.~Brdar, P.~Schwaller, and A.~Segner, {\it {Shining Light on
  the Scotogenic Model: Interplay of Colliders and Cosmology}},  {\em JHEP}
  {\bf 09} (2020) 136, [\href{http://xxx.lanl.gov/abs/1912.0821}{{\tt
  arXiv:1912.0821}}].

\bibitem{Esch2018}
S.~Esch, M.~Klasen, and C.~E. Yaguna, {\it {A singlet doublet dark matter model
  with radiative neutrino masses}},  {\em JHEP} {\bf 10} (2018) 055,
  [\href{http://xxx.lanl.gov/abs/1804.0338}{{\tt arXiv:1804.0338}}].

\bibitem{Sarazin:2021nwo}
M.~Sarazin, J.~Bernigaud, and B.~Herrmann, {\it {Dark matter and lepton flavour
  phenomenology in a singlet-doublet scotogenic model}},  {\em JHEP} {\bf 12}
  (2021) 116, [\href{http://xxx.lanl.gov/abs/2107.0461}{{\tt
  arXiv:2107.0461}}].

\bibitem{Alvarez:2023dzz}
A.~Alvarez, A.~Banik, R.~Cepedello, B.~Herrmann, W.~Porod, M.~Sarazin, and
  M.~Schnelke, {\it {Accommodating muon (g \ensuremath{-} 2) and leptogenesis
  in a scotogenic model}},  {\em JHEP} {\bf 06} (2023) 163,
  [\href{http://xxx.lanl.gov/abs/2301.0848}{{\tt arXiv:2301.0848}}].

\bibitem{Betancur:2017dhy}
A.~Betancur, R.~Longas, and O.~Zapata, {\it {Doublet-triplet dark matter with
  neutrino masses}},  {\em Phys. Rev. D} {\bf 96} (2017), no.~3 035011,
  [\href{http://xxx.lanl.gov/abs/1704.0116}{{\tt arXiv:1704.0116}}].

\bibitem{Betancur:2018xtj}
A.~Betancur and O.~Zapata, {\it {Phenomenology of doublet-triplet fermionic
  dark matter in nonstandard cosmology and multicomponent dark sectors}},  {\em
  Phys. Rev. D} {\bf 98} (2018), no.~9 095003,
  [\href{http://xxx.lanl.gov/abs/1809.0499}{{\tt arXiv:1809.0499}}].

\bibitem{Restrepo2013}
D.~Restrepo, O.~Zapata, and C.~E. Yaguna, {\it {Models with radiative neutrino
  masses and viable dark matter candidates}},  {\em JHEP} {\bf 11} (2013) 011,
  [\href{http://xxx.lanl.gov/abs/1308.3655}{{\tt arXiv:1308.3655}}].

\bibitem{ParticleDataGroup:2022pth}
{\bf Particle Data Group} , R.~L. Workman et~al., {\it {Review of Particle
  Physics}},  {\em PTEP} {\bf 2022} (2022) 083C01.

\bibitem{Pontecorvo1957b}
B.~Pontecorvo, {\it {Inverse beta processes and nonconservation of lepton
  charge}},  {\em Sov. Phys. JETP} {\bf 7} (1958) 172--173. [Zh. Eksp. Teor.
  Fiz.34,247(1957)].

\bibitem{Maki1962}
Z.~Maki, M.~Nakagawa, and S.~Sakata, {\it {Remarks on the unified model of
  elementary particles}},  {\em Prog. Theor. Phys.} {\bf 28} (1962) 870--880.

\bibitem{Markov1971}
A.~A. Markov, {\em Extension of the limit theorems of probability theory to a
  sum of variables connected in a chain}.
\newblock reprinted in Appendix B of: R. Howard, {\it Dynamic Probabilistic
  Systems, volume 1: Markov Chains}, John Wiley and Sons, 1971.

\bibitem{Metropolis1953}
N.~Metropolis, A.~W. Rosenbluth, M.~N. Rosenbluth, A.~H. Teller, and E.~Teller,
  {\it Equation of state calculations by fast computing machines},  {\em J.
  Chem. Phys.} {\bf 21} (1953) 1087--1092.

\bibitem{Hastings1970}
W.~K. Hastings, {\it {Monte Carlo Sampling Methods Using Markov Chains and
  Their Applications}},  {\em Biometrika} {\bf 57} (1970) 97--109.

\bibitem{CasasIbarra2001}
J.~A. Casas and A.~Ibarra, {\it {Oscillating neutrinos and $\mu \to e,
  \gamma$}},  {\em Nucl. Phys. B} {\bf 618} (2001) 171--204,
  [\href{http://xxx.lanl.gov/abs/hep-ph/0103065}{{\tt hep-ph/0103065}}].

\bibitem{SPheno2003}
W.~Porod, {\it {SPheno, a program for calculating supersymmetric spectra, SUSY
  particle decays and SUSY particle production at $e^+ e^-$ colliders}},  {\em
  Comput. Phys. Commun.} {\bf 153} (2003) 275--315,
  [\href{http://xxx.lanl.gov/abs/hep-ph/0301101}{{\tt hep-ph/0301101}}].

\bibitem{SPheno2012}
W.~Porod and F.~Staub, {\it {SPheno 3.1: Extensions including flavour,
  CP-phases and models beyond the MSSM}},  {\em Comput. Phys. Commun.} {\bf
  183} (2012) 2458--2469, [\href{http://xxx.lanl.gov/abs/1104.1573}{{\tt
  arXiv:1104.1573}}].

\bibitem{SARAH2010}
F.~Staub, {\it {From Superpotential to Model Files for FeynArts and
  CalcHep/CompHep}},  {\em Comput. Phys. Commun.} {\bf 181} (2010) 1077--1086,
  [\href{http://xxx.lanl.gov/abs/0909.2863}{{\tt arXiv:0909.2863}}].

\bibitem{SARAH2011}
F.~Staub, {\it {Automatic Calculation of supersymmetric Renormalization Group
  Equations and Self Energies}},  {\em Comput. Phys. Commun.} {\bf 182} (2011)
  808--833, [\href{http://xxx.lanl.gov/abs/1002.0840}{{\tt arXiv:1002.0840}}].

\bibitem{SARAH2013}
F.~Staub, {\it {SARAH 3.2: Dirac Gauginos, UFO output, and more}},  {\em
  Comput. Phys. Commun.} {\bf 184} (2013) 1792--1809,
  [\href{http://xxx.lanl.gov/abs/1207.0906}{{\tt arXiv:1207.0906}}].

\bibitem{SARAH2014}
F.~Staub, {\it {SARAH 4: A tool for (not only SUSY) model builders}},  {\em
  Comput. Phys. Commun.} {\bf 185} (2014) 1773--1790,
  [\href{http://xxx.lanl.gov/abs/1309.7223}{{\tt arXiv:1309.7223}}].

\bibitem{Martin:1997ns}
S.~P. Martin, {\it {A Supersymmetry primer}},  {\em Adv. Ser. Direct. High
  Energy Phys.} {\bf 18} (1998) 1--98,
  [\href{http://xxx.lanl.gov/abs/hep-ph/9709356}{{\tt hep-ph/9709356}}].

\bibitem{MO2004}
G.~B\'elanger, F.~Boudjema, A.~Pukhov, and A.~Semenov, {\it {micrOMEGAs:
  Version 1.3}},  {\em Comput. Phys. Commun.} {\bf 174} (2006) 577--604,
  [\href{http://xxx.lanl.gov/abs/hep-ph/0405253}{{\tt hep-ph/0405253}}].

\bibitem{Harz:2023llw}
J.~Harz, B.~Herrmann, M.~Klasen, K.~Kova\v{r}\'\i{}k, and L.~P. Wiggering, {\it
  {Precision predictions for dark matter with DM@NLO in the MSSM}},  {\em Eur.
  Phys. J. C} {\bf 84} (2024), no.~4 342,
  [\href{http://xxx.lanl.gov/abs/2312.1720}{{\tt arXiv:2312.1720}}].

\bibitem{LZ:2022lsv}
{\bf LZ} , J.~Aalbers et~al., {\it {First Dark Matter Search Results from the
  LUX-ZEPLIN (LZ) Experiment}},  {\em Phys. Rev. Lett.} {\bf 131} (2023), no.~4
  041002, [\href{http://xxx.lanl.gov/abs/2207.0376}{{\tt arXiv:2207.0376}}].

\bibitem{MO2001}
G.~B\'elanger, F.~Boudjema, A.~Pukhov, and A.~Semenov, {\it {MicrOMEGAs: A
  Program for calculating the relic density in the MSSM}},  {\em Comput. Phys.
  Commun.} {\bf 149} (2002) 103--120,
  [\href{http://xxx.lanl.gov/abs/hep-ph/0112278}{{\tt hep-ph/0112278}}].

\bibitem{MO2007a}
G.~B\'elanger, F.~Boudjema, A.~Pukhov, and A.~Semenov, {\it {MicrOMEGAs 2.0: A
  Program to calculate the relic density of dark matter in a generic model}},
  {\em Comput. Phys. Commun.} {\bf 176} (2007) 367--382,
  [\href{http://xxx.lanl.gov/abs/hep-ph/0607059}{{\tt hep-ph/0607059}}].

\bibitem{MO2007b}
G.~B\'elanger, F.~Boudjema, A.~Pukhov, and A.~Semenov, {\it {micrOMEGAs 2.0.7:
  A program to calculate the relic density of dark matter in a generic model}},
   {\em Comput. Phys. Commun.} {\bf 177} (2007) 894--895.

\bibitem{MO2013}
G.~B\'elanger, F.~Boudjema, A.~Pukhov, and A.~Semenov, {\it {micrOMEGAs 3: A
  program for calculating dark matter observables}},  {\em Comput. Phys.
  Commun.} {\bf 185} (2014) 960--985,
  [\href{http://xxx.lanl.gov/abs/1305.0237}{{\tt arXiv:1305.0237}}].

\bibitem{MO2018}
G.~B\'elanger, F.~Boudjema, A.~Goudelis, A.~Pukhov, and B.~Zaldivar, {\it
  {micrOMEGAs5.0 : Freeze-in}},  {\em Comput. Phys. Commun.} {\bf 231} (2018)
  173--186, [\href{http://xxx.lanl.gov/abs/1801.0350}{{\tt arXiv:1801.0350}}].

\bibitem{FromPDG_TheMEG:2016wtm}
{\bf MEG} , A.~M. Baldini et~al., {\it {Search for the lepton flavour violating
  decay $\mu ^+ \rightarrow \mathrm {e}^+ \gamma $ with the full dataset of the
  MEG experiment}},  {\em Eur. Phys. J. C} {\bf 76} (2016), no.~8 434,
  [\href{http://xxx.lanl.gov/abs/1605.0508}{{\tt arXiv:1605.0508}}].

\bibitem{Blondel:2013ia}
A.~Blondel et~al., {\it {Research Proposal for an Experiment to Search for the
  Decay $\mu \to eee$}},  \href{http://xxx.lanl.gov/abs/1301.6113}{{\tt
  arXiv:1301.6113}}.

\bibitem{SINDRUMII:2006dvw}
{\bf SINDRUM II} , W.~H. Bertl et~al., {\it {A Search for muon to electron
  conversion in muonic gold}},  {\em Eur. Phys. J. C} {\bf 47} (2006) 337--346.

\bibitem{Rule:2024kjo}
E.~Rule, {\it {Nucleon-level Effective Theory of $\mu \to e$ Conversion}},
  {\em PoS} {\bf CD2021} (2024) 099.

\bibitem{Calibbi:2017uvl}
L.~Calibbi and G.~Signorelli, {\it {Charged Lepton Flavour Violation: An
  Experimental and Theoretical Introduction}},  {\em Riv. Nuovo Cim.} {\bf 41}
  (2018), no.~2 71--174, [\href{http://xxx.lanl.gov/abs/1709.0029}{{\tt
  arXiv:1709.0029}}].

\bibitem{Natori:2014yba}
{\bf DeeMe} , H.~Natori, {\it {DeeMe experiment - An experimental search for a
  mu-e conversion reaction at J-PARC MLF}},  {\em Nucl. Phys. B Proc. Suppl.}
  {\bf 248-250} (2014) 52--57.

\bibitem{Meucci:2022qbh}
{\bf MEG II} , M.~Meucci, {\it {MEG II experiment status and prospect}},  {\em
  PoS} {\bf NuFact2021} (2022) 120,
  [\href{http://xxx.lanl.gov/abs/2201.0820}{{\tt arXiv:2201.0820}}].

\bibitem{COMET2024}
A.~Artikov, V.~Baranov, A.~Boikov, D.~Chokheli, Y.~Davydov, V.~Glagolev,
  A.~Simonenko, Z.~Tsamalaidze, I.~Vasilyev, and I.~Zimin, {\it High efficiency
  muon registration system based on scintillator strips},  {\em Nuclear
  Instruments and Methods in Physics Research Section A: Accelerators,
  Spectrometers, Detectors and Associated Equipment} {\bf 1064} (2024) 169436.

\bibitem{Moritsu:2022lem}
{\bf COMET} , M.~Moritsu, {\it {Search for Muon-to-Electron Conversion with the
  COMET Experiment \textdagger{}}},  {\em Universe} {\bf 8} (2022), no.~4 196,
  [\href{http://xxx.lanl.gov/abs/2203.0636}{{\tt arXiv:2203.0636}}].

\bibitem{deGouvea:2013zba}
A.~de~Gouvea and P.~Vogel, {\it {Lepton Flavor and Number Conservation, and
  Physics Beyond the Standard Model}},  {\em Prog. Part. Nucl. Phys.} {\bf 71}
  (2013) 75--92, [\href{http://xxx.lanl.gov/abs/1303.4097}{{\tt
  arXiv:1303.4097}}].

\bibitem{NuFit2019}
{Nu-Fit~4.1}, {\em Three-neutrino fit based on data available in July 2019},
  2019.
\newblock \url{http://www.nu-fit.org}.

\bibitem{NuFit2018}
I.~Esteban, M.~C. Gonzalez-Garcia, A.~Hernandez-Cabezudo, M.~Maltoni, and
  T.~Schwetz, {\it {Global analysis of three-flavour neutrino oscillations:
  synergies and tensions in the determination of $\theta_{23}$, $\delta_{CP}$,
  and the mass ordering}},  {\em JHEP} {\bf 01} (2019) 106,
  [\href{http://xxx.lanl.gov/abs/1811.0548}{{\tt arXiv:1811.0548}}].

\bibitem{CalcHep_Belyaev:2012qa}
A.~Belyaev, N.~D. Christensen, and A.~Pukhov, {\it {CalcHEP 3.4 for collider
  physics within and beyond the Standard Model}},  {\em Comput. Phys. Commun.}
  {\bf 184} (2013) 1729--1769, [\href{http://xxx.lanl.gov/abs/1207.6082}{{\tt
  arXiv:1207.6082}}].

\bibitem{DARWIN:2016hyl}
{\bf DARWIN} , J.~Aalbers et~al., {\it {DARWIN: towards the ultimate dark
  matter detector}},  {\em JCAP} {\bf 11} (2016) 017,
  [\href{http://xxx.lanl.gov/abs/1606.0700}{{\tt arXiv:1606.0700}}].

\bibitem{XENONnT2020}
{\bf XENON} , E.~Aprile et~al., {\it {Projected WIMP sensitivity of the XENONnT
  dark matter experiment}},  {\em JCAP} {\bf 11} (2020) 031,
  [\href{http://xxx.lanl.gov/abs/2007.0879}{{\tt arXiv:2007.0879}}].

\bibitem{Armand:2022sjf}
C.~Armand and B.~Herrmann, {\it {Dark matter indirect detection limits from
  complete annihilation patterns}},  {\em JCAP} {\bf 11} (2022) 055,
  [\href{http://xxx.lanl.gov/abs/2210.0122}{{\tt arXiv:2210.0122}}].

\bibitem{Boruah:2021qlf}
B.~B. Boruah, L.~Sarma, and M.~K. Das, {\it {Lepton flavor violation and
  leptogenesis in discrete flavor symmetric scotogenic model}},
  \href{http://xxx.lanl.gov/abs/2103.0529}{{\tt arXiv:2103.0529}}.

\bibitem{g2Exp}
{\bf Muon $g\ensuremath{-}2$ Collaboration} , D.~P. Aguillard et~al., {\it
  Detailed report on the measurement of the positive muon anomalous magnetic
  moment to 0.20 ppm},  {\em Phys. Rev. D} {\bf 110} (Aug, 2024) 032009.

\bibitem{g-2SM}
A.~Keshavarzi, K.~S. Khaw, and T.~Yoshioka, {\it Muon $g-2$: A review},  {\em
  Nuclear Physics B} {\bf 975} (2022) 115675.

\bibitem{SLHA1}
P.~Z. Skands et~al., {\it {SUSY Les Houches accord: Interfacing SUSY spectrum
  calculators, decay packages, and event generators}},  {\em JHEP} {\bf 07}
  (2004) 036, [\href{http://xxx.lanl.gov/abs/hep-ph/0311123}{{\tt
  hep-ph/0311123}}].

\bibitem{SLHA2}
B.~C. Allanach et~al., {\it {SUSY Les Houches Accord 2}},  {\em Comput. Phys.
  Commun.} {\bf 180} (2009) 8--25,
  [\href{http://xxx.lanl.gov/abs/0801.0045}{{\tt arXiv:0801.0045}}].

\bibitem{MatPlotLib}
J.~D. Hunter, {\it Matplotlib: A 2d graphics environment},  {\em Computing In
  Science \& Engineering} {\bf 9} (2007), no.~3 90--95.

\end{thebibliography}\endgroup
